\documentclass[a4paper,11pt]{article}

\pdfoutput=0  % for eps graphics files: this tex file uses dvips and ps2pdf

\usepackage[T1]{fontenc}   
\usepackage[applemac]{inputenc}
\usepackage{colortbl}           % for color rules  
\usepackage{booktabs}           % for good tables  
\usepackage{longtable}
\usepackage{array}              % for mathematical tables
\usepackage{caption}

\usepackage{times}
\usepackage[english]{babel}
\usepackage{amsmath,amssymb,url}
\renewcommand\url{\begingroup \urlstyle{same}\Url}

\usepackage[dvips]{graphicx}
\DeclareGraphicsExtensions{.eps}

\normalsize

\newcommand{\PBS}[1]{\let\temp=\\#1\let\\=\temp} 
\let\s\sigma

\def\csmatl#1{\left(\begin{array}{{#1}}}
\def\csmatr{\end{array}\right)}
\def\csd#1#2{\leavevmode\hbox{${#1}$}\,\hbox{${\mathrm{#2}}$}}
\def\csbul{\leavevmode{\rule[0.14em]{0.24em}{0.24em}}\kern 0.33em}
\def\csbullet{\leavevmode{\rule[0.14em]{0.24em}{0.24em}}\kern 0.33em}
\def\cp{\;.}

\def\cv{\;,\;}

\def\figureref#1{{{Figure~\ref{#1}}}}
\def\tableref#1{{{Table~\ref{#1}}}}
  
\def\seepage#1{\relax}

\def\challengedif#1{\relax}

\makeatletter

\def\iin{\@ifnextchar[{\iinhelpone}{\iinhelptwo}}
\def\iinhelptwo#1{\iinhelpone[#1]{#1}}
\def\iinhelpone[#1]#2{\leavevmode\index{#1}#2}

\def\ii{\@ifnextchar[{\iihelpone}{\iihelptwo}}
\def\iihelptwo#1{\iihelpone[#1]{#1}}
\def\iihelpone[#1]#2{\leavevmode\indexs{#1}\emph{#2}}
\def\indexs#1{\index{#1|bbb}}

\def\csepsf{\@ifnextchar[{\@cssepsf}{\@cssepsf[t]}} 
\def\@cssepsf[#1]#2#3#4{% 
  \begin{figure}[tp]%
  \hbox to \textwidth{\hss{\includegraphics[#3]{#2}}\hss}%
  \caption{#4}%
  \protect\label{#2}%
\end{figure}}%
      
\def\csrhd{\hbox to 0 pt{\hss\hbox to 13 pt{$\rhd$\hss}}}

\makeatother

\begin{document}

\title{Deducing the three gauge interactions  \\ 
from the three Reidemeister moves \\  \ }

\author{Christoph Schiller\\
Physik Department T37, Technical University Munich,\\
James-Frank-Str., 85748 Garching, Germany\\
{christoph.schiller@ph.tum.de}}

\date{20 October 2010}

\maketitle\normalfont\normalsize

\abstract{\noindent Possibly the first argument for the origin of the three
observed gauge groups and thus for the origin of the three non-gravitational
interactions is presented.  The argument is based on a proposal for the final
theory that models nature at Planck scales as a collection of featureless
strands that fluctuate in three dimensions.  This approach models vacuum as
untangled strands, particles as tangles of strands, and Planck units as
crossing switches.

Modelling vacuum as untangled strands implies the field equations of general
relativity, when applying an argument from 1995 to the thermodynamics of
such strands.  Modelling fermions as tangles of two or more strands allows to
define wave functions as time-averages of oriented strand crossing density.
Using an argument from 1980, this allows to deduce the Dirac equation.

When fermions are modelled as tangled strands, gauge interactions appear
naturally as deformation of tangle cores.  The three possible types of
observable core deformations are given by the three Reidemeister moves.  They
naturally lead to a U(1), a broken and parity-violating SU(2), and a SU(3)
gauge group.  The corresponding Lagrangians also appear naturally.

The strand model is unique, is unmodifiable, is consistent with all known
data, and makes numerous testable predictions, including the absence of other
interactions, of grand unification, of supersymmetry and of higher dimensions.
A method for calculating coupling constants seems to appear naturally.

\bigskip
\bigskip

\noindent Keywords:  gauge interactions, strand model, standard model, coupling
constants

\medskip

\noindent PACS numbers: 
12.10.-g, % unified field theories and models
12.60.Rc  % composite models
}

\newpage\normalfont\normalsize

%-----------------------------------------------------------------------------
% Feb 2009
\section{Planck units, strands and unification}

Physics as we know it today, i.e., quantum field theory and general
relativity, is a low energy version of physics at the Planck scale.  Effects
of the Planck scale are known to be most evident on horizons, especially on
black hole horizons.  A basic result of twentieth-century physics is that at
horizons, vacuum and particles mix.  Therefore, we can guess that particles
and vacuum are made of common constituents.  In addition, the surface
dependence of black hole entropy tells us that these constituents are not
point-like, but extended.

Three questions ensue.  First, what is the simplest description of nature that
contains these results?  Second, can the standard model of elementary
particles be deduced from such a description?  Third, is such a description
unified?  We shall argue that the answer to the first question are fluctuating
{featureless strands} in three spatial dimensions, and that the answers to the
second and third question are affirmative.

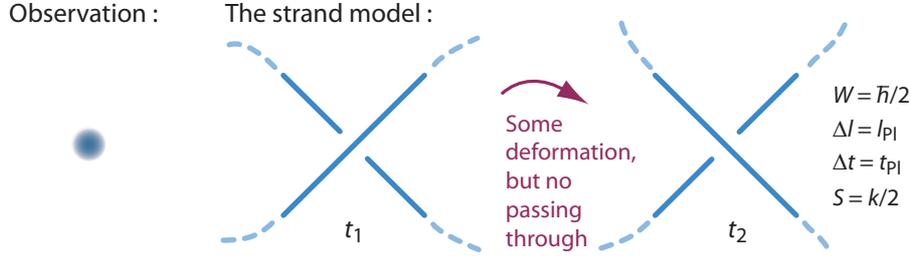
\csepsf{i-eventatomp4}{scale=1}{The simplest observation, a `point-like' event
in three spatial dimensions, and its associated strand model.}

It is known that the observer invariance of Planck units, in particular the
invariance of the maximum speed $c$, the invariance of the quantum of action
$\hbar$ and the invariance of the Planck force $c^4/4G$, are sufficient to
deduce the Dirac equation of quantum theory \cite{bpr,cs3} and Einstein's
field equations of general relativity \cite{jacobgg,cs2a}. 

In contrast to a widespread opinion, a model of physics at the Planck scale
therefore does not need to start from the Dirac equation or from the Einstein
equations; it is sufficient to start from the invariance of the Planck units.
As long as a unified model leaves the Planck units invariant for all
observers, the Dirac equation and the Einstein equations are automatically
satisfied.

Probably the simplest model that allows the transition from Planck units to
quantum field theory and to general relativity as intuitive as possible is a
model based on \emph{featureless strands}.  Strands, not points, are assumed
to be the fundamental constituents of matter, radiation and vacuum.  In other
words,

\begin{quotation}
    \noindent \csrhd {\textbf{Nature} is assumed to be made of fluctuating
    strands in three spatial dimensions.}
\end{quotation}

\noindent To describe observations, the strand model uses one new basic
postulate:

\begin{quotation}
    \noindent \csrhd {Each \textbf{Planck unit} and each \textbf{event} is the
    switch of a crossing between two strand segments.}
\end{quotation}

\noindent This definition of an event as a crossing switch is illustrated in
\figureref{i-eventatomp4}.  Every event is characterized by Planck's quantum
of action $\hbar$, the Planck time $t_{\rm Pl}=\sqrt{G\hbar/c^5}$, the Planck
length $l_{\rm Pl}=\sqrt{G\hbar/c^3}$, and the Planck entropy, i.e., the
Boltzmann constant $k$.  More precisely, the process shown in
\figureref{i-eventatomp4} corresponds to an action $\hbar/2$, while $\hbar$
corresponds to a full turn.  (For any specific shape change, the number of
crossing switches is observer-dependent; this is related to Lorentz and gauge
transformations.)  The basic postulate thus declares that events are not
points on manifolds, but (observable) crossing switches of (unobservable)
strands.  Events are thus only point-like in the approximation that the Planck
length is negligibly small.

The strands are \emph{featureless}: they have no mass, no tension, no
branches, no fixed length, no diameter, no ends and cannot be cut.  Strands
have no observable property at all: strands are unobservable.  Only crossing
switches are observable.
% Strands cannot cross each other.

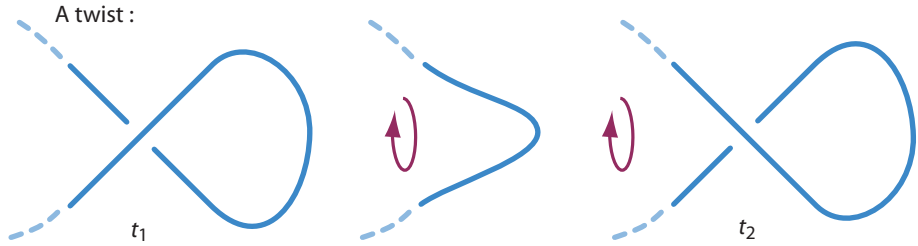
\csepsf{i-switch-examples}{scale=1}{An example of strand deformation
leading to a crossing switch.}

Above all, strands are \emph{impenetrable}; realizing a crossing switch thus
always requires the motion of strand segments \emph{around} each other.  A
simple example of deformation leading to a crossing switch is shown in
\figureref{i-switch-examples}.

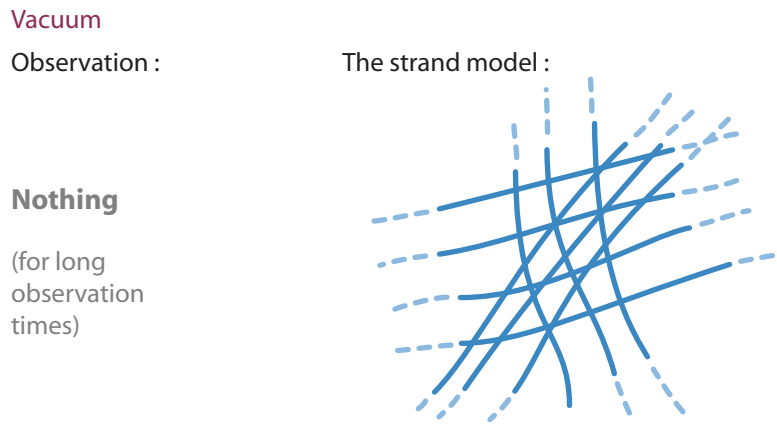
\csepsf{i-vacuum}{scale=1}{A schematic illustration of the strand
model for the vacuum.}

The strand model asserts that matter, radiation and vacuum are all built from
fluctuating strands.  In particular, 

\begin{quotation}
    \noindent \csrhd {\textbf{Vacuum} is made of fluctuating
    unknotted and untangled strands.}
\end{quotation}

\noindent Flat vacuum shows, averaged over time, no knots, no tangles and no
other crossing switches, so that it is observed to be empty of matter or
radiation, as illustrated in \figureref{i-vacuum}.  Independently of the
strand model, in a recent exploration of the small scale structure of
space-time from various different research perspectives in general relativity,
{Steven Carlip} also comes to the conclusion that all these perspectives
suggest the common idea that ``space at a fixed time is thus threaded by
rapidly fluctuating lines'' \cite{cup}.

\begin{quotation}
    \noindent \csrhd \textbf{Continuous,
three-dimensional background} space is introduced by the observer, in order to
describe observations.
\end{quotation}

\noindent Describing nature without a background seems impossible.  To be
self-consistent, the background must arise, as it does here, from the
time-average of the vacuum strands.  In short, the strand model makes use both
of discrete strands and of a continuous background.  Curvature and horizons
have a natural description in terms of strands; the model proposes that
strands describe quantum geometry at Planck scales.  Indeed, exploring the
thermodynamics of strands yields the field equations of general relativity, a
natural model of event horizons as strand weaves, and black hole entropy
\cite{jacobgg,cs2}.  Universal gravity emerges in the exactly the same way as
explained recently \cite{ev,pad}.  These connections are not topic of the
present paper, however.

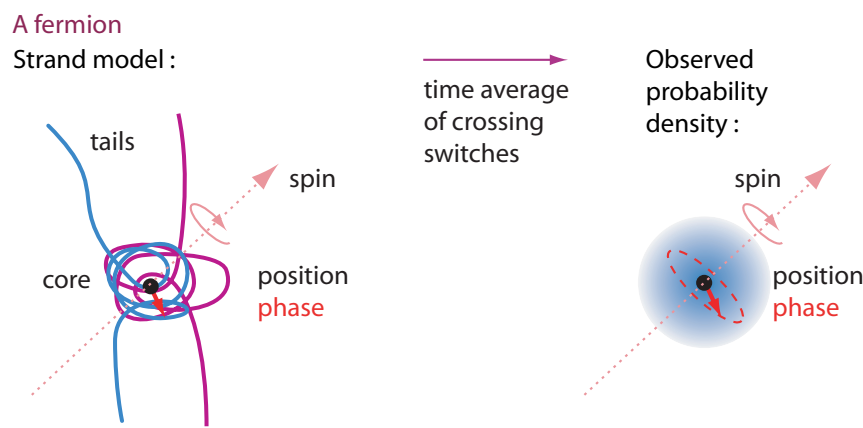
\csepsf{i-tanglerotp4}{scale=1}{A massive free spin 1/2 \emph{fermion} built
of tangled strands with its core and tails (left) and the corresponding
probability cloud that results from averaging its crossing switch distribution
over time (right), showing the corresponding particle position and phase.}

\begin{quotation}
    \noindent \csrhd \textbf{Particles} are tangles of strands.
\end{quotation}

\noindent As shown below, this definition of particles can yield spin 1/2 or
spin 1 behaviour for elementary particles, depending on the tangle details.
In particular, tangles lead to a model of {fermions} illustrated in
\figureref{i-tanglerotp4} and \figureref{i-belttrick}.  This model of fermions
is known to allow deducing the Dirac equation \cite{bpr,cs3}; the argument
will be summarized below.

In summary, the strand model appears to make it possible to describe nature,
and in particular vacuum, matter and radiation, with the help of fluctuating
strands in a three-dimensional background defined by the observer.
\tableref{physsystang} lists the main correspondences between physical systems
and tangles.  In all physical systems, the shape fluctuations of tangles lead
to crossing switches and thus, indirectly, to the usual evolution of matter,
radiation and vacuum curvature.  Crossing switches are used to define the
fundamental physical units, and thus all physical observables.

Before even exploring the physical consequences, two issues arise: (1) Where
do fluctuations come from?  (2) What is their influence on the dynamics?  The
strand model argues that strand fluctuations arise automatically, whenever a
continuous, three-dimensional background space-time is introduced by the
observer, and that the fluctuations of a particular piece of strand are due to
all other pieces of strands in the universe.  The model also argues that the
fluctuations have precisely the behaviour that allows introducing a
background: the fluctuations are homogeneous and isotropic.  In particular,
whenever two strands approach each other, the fluctuations of the two strands
become correlated due to their impenetrability, but the embedding in a
three-dimensional background remains possible (on a local scale).  Whenever a
background can be introduced (i.e., whenever fluctuations are, on average,
locally homogeneous and isotropic), Einstein's field equations and Dirac's
equation are not sensitive to any assumed detailed properties or dynamics of
the fluctuations, as long as the fluctuations lead to the usual space-time
symmetries \cite{cs2, cs3}.

% \end{document}
% \endinput

%-----------------------------------------------------------------------------
% \subsubsubsubsubsubsubsubsection{Table of physics-tangle correspondences}
%
\begin{table}[t]
\small
\centering
\caption{Correspondences between physical systems and mathematical
tangles.\label{physsystang}}
\begin{tabular*}{\textwidth}{@{} >{\PBS\raggedright} p{46mm} 
    !{\extracolsep{\fill}} >{\PBS\raggedright} p{51mm} !{\extracolsep{\fill}}
    >{\PBS\raggedright} p{43mm}@{\hspace{0em}}}
\toprule
\textsc{Physical system} & \textsc{Strands} & \textsc{Tangle type} \\[0.5mm]
\midrule
Vacuum and dark energy & many infinite unknotted strands & unlinked \\
Elementary vector boson (radiation) & one infinite strand & trivial or knotted 
curve\\
Elementary fermion (matter) & two or three infinite strands & braided,
rational or prime tangle\\
Graviton & two infinite twisted strands & rational tangle \\
Gravity wave & many infinite twisted strands & many rational tangles\\
Horizon & many woven infinite strands & weave-like tangle\\
\bottomrule
\end{tabular*}
\bigskip
\end{table}
%-----------------------------------------------------------------------------

%-----------------------------------------------------------------------------
% Sep 2006
\subsection{Appearance and unification of interactions}

We will argue that the definition of an event as a crossing switch of strands
yields a model for the three gauge interactions.  A separate discussion
specifies the tangle structure for each elementary particle \cite{cs5}.  For
the following discussion, we assume \emph{flat} space-time.

To show the natural appearance of exactly three gauge interactions from the
basic postulate, we will use three older results and a new one: 

\medskip

\noindent (1) tangles of
strands allow to define wave functions and reproduce the Dirac equation
\cite{bpr,cs3}, 

\noindent (2) shape deformations are equivalent to gauge theory
\cite{berry,shapwil,squarecat}, 

\noindent (3) all observable tangle deformations can be
reduced to only three types \cite{reidem,lou}.  

\medskip

\noindent As we will see, these three results allow to deduce the U(1) and the
broken SU(2) gauge symmetries, including the corresponding Lagrangians.  With
the new result that

\medskip 

\noindent  (4) the group SU(3) appears in an
expanded belt trick with three belts, 

\medskip
\noindent   also the strong interaction Lagrangian
is recovered.

We thus aim to deduce the main aspects of the Lagrangian of the standard model
from strands.  To do this, we first recall briefly how the Dirac equation for
\emph{free}, non-interacting, elementary spin 1/2 particles and its Lagrangian
are deduced.  We start with the spin aspects.

%-----------------------------------------------------------------------------
% Sep 2006, Nov 2006
\subsection{Spin and statistics}

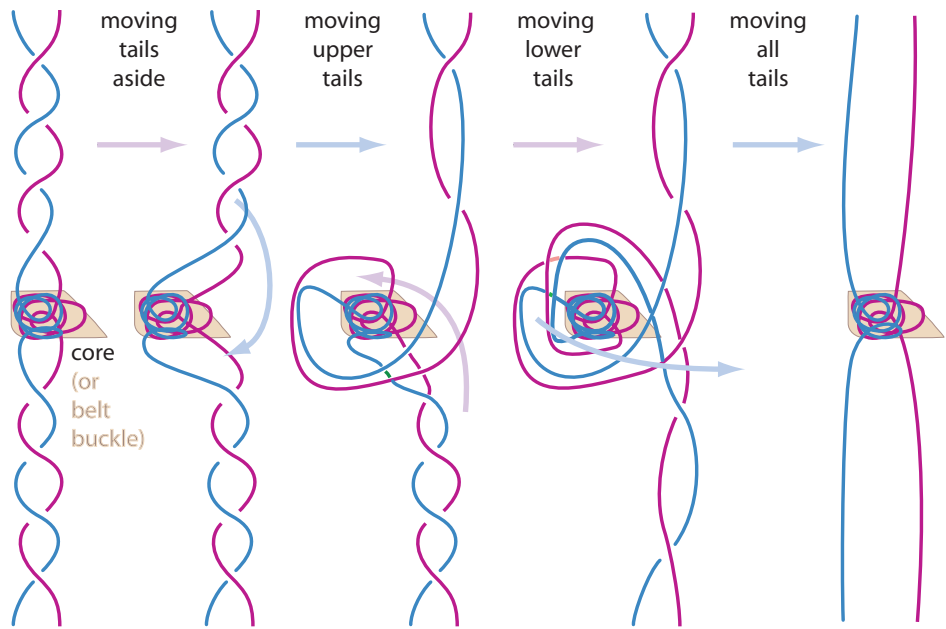
\csepsf{i-belttrick}{scale=1}{The \emph{belt trick} or \emph{scissor trick} or
\emph{string trick}: rotations by 4 $\pi$ of an object with three or more
tails are equivalent to no rotation -- allowing a suspended object,
such as a belt buckle or a tangle core, to rotate for ever.  Note that there
are \emph{two} ways to perform the belt trick after a rotation by 4 $\pi$ is
completed, as is best illustrated by an internet animation
\protect\cite{egan}.}

\noindent It is known since many decades that so-called \emph{belt trick} --
also called \emph{scissor trick} or \emph{string trick} -- illustrated in
\figureref{i-belttrick}, can be used, together with its variations, to model
spin 1/2 behaviour.  It is also well-known that fluctuating strands with tails
reaching the `border' of space reproduce the spin-statistics theorem for
bosons and fermions, depending on the tangle details \cite{cs3}.

The belt trick implies that for fermions made of \emph{two or more} tangled
strands, and thus with four or more tails to the `border', as shown in
\figureref{i-tanglerotp4} and \figureref{i-belttrick}, a rotation by $4 \pi$
of the tangle core -- thus a rotation by \emph{two} full turns -- brings back
such a tangle to the original state.  In short, the belt trick allows a object
that is attached by strands to spatial infinity to rotate indefinitely.  The
same is valid for a tangle core.

In addition, after exchanging two tangle cores {twice}, tail fluctuations
alone can return the situation to the original state.  Cores made of two or
more tangled strands thus behave both under rotations and under exchange like
spin 1/2 particles.

For cores made of \emph{one} strand -- thus with two tails to the border -- a
rotation by $2 \pi$ restores the original state.  Such a core, shown in
\figureref{i-ee-boson}, behaves like a spin 1 particle, thus like a boson.

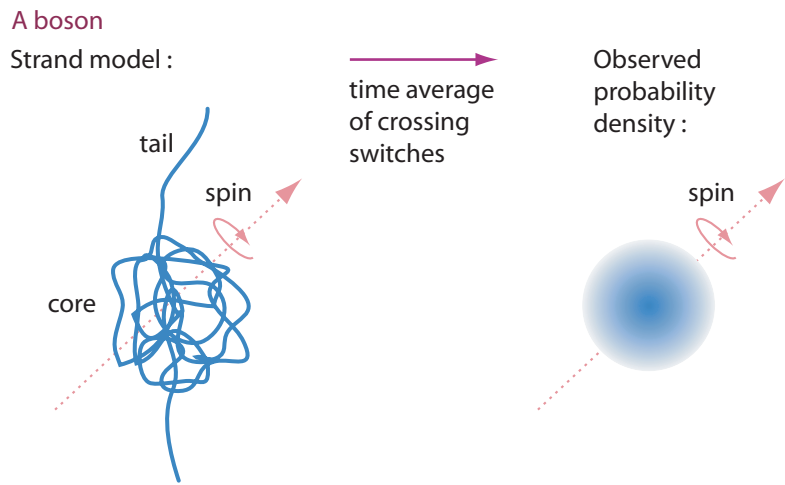
\csepsf{i-ee-boson}{scale=1}{A massive spin 1 \emph{boson} in the strand
model (left) and the observed probability density when averaging its crossing
switches over long time scales (right).}

As shown in \figureref{i-vacuum}, in the strand model, the vacuum is modelled
as a collection of untangled strands.  A strand model for the graviton,
invariant under rotations by $\pi$ and thus with spin 2, has been introduced
in the discussion of general relativity \cite{cs2}.

In short, the spin-statistics connection for all elementary particles can be
reproduced by the strand model.  All evolution and all particle reactions
\emph{conserve} spin, because all interactions conserve the number of strands
and tails, as will be detailed below.  The strand model also implies that spin
values are always an \emph{integer multiple of 1/2}.  In summary, the strand
model reproduces spin in all its observed details \cite{cs3}.  In particular,
tangles of two or more strands reproduce spin 1/2 particles.  

%-----------------------------------------------------------------------------
% Nov 2009
\subsection{Wave functions for fermions}

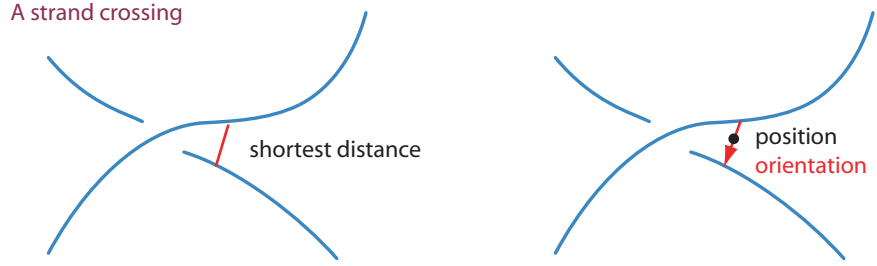
\csepsf{i-strand-crossing}{scale=1}{The definition of a crossing,  its
position and its orientation.}

We now summarize the definition of wave functions using strands.  Given a
fluctuating tangle of several strands, we define:

\begin{quotation}
    \noindent \csrhd The \textbf{wave function} of a system described by a
    tangle is the \emph{time average} of the positions and the orientations of
    its crossings (and thus \emph{not} of its crossing \emph{switches}).
\end{quotation}

\noindent Each crossing of two strand segments in a three-dimensional
background defines a line of shortest distance, as shown schematically in
\figureref{i-strand-crossing}.  The centre of this line defines a position and
an orientation for each crossing.  For the definition of the wave function,
the \textbf{time average} of crossing positions and orientations is taken over
the typical time resolution of the observer.  This is a time interval that is
much \emph{longer} than the Planck time, but also much \emph{shorter} than the
typical evolution time of the system.  The time resolution is what the
observer calls an `instant' of time.  Typically, the resolution will be
\csd{10^{-25}}{s} or longer; the typical averaging will thus be over all times
between \csd{10^{-43}}{s}, the Planck time, and \csd{10^{-25}}{s} or more.

The wave function can thus be called the `oriented crossing density'.  As
such, it is a continuous function of space.  The wave function thus captures
the local time average of all possible tangle fluctuations.  For a tangle with
a few crossings, \figureref{i-tanglerotp4} illustrates the idea.  However, the
figure does not show the wave function itself, but its probability density.
In fact, the probability density is the (square root of the) crossing
\emph{position} density, whereas the wave function is a density that describes
\emph{both position and orientation} of crossings.  The strand model thus
visualizes the idea of wave function with an \emph{oriented cloud}.  We also
mention that the Hilbert structure is easily deduced from the strand
definition of wave functions \cite{cs3}.

For particles with spin 1/2, not only the time-average of crossing density is
a function of space; also the time-average of the spin axis orientation is.
One scalar, $\rho(x,t)$, is needed to describe the crossing density, and three
Euler angles, $\alpha$, $\beta$ and $\gamma$, are needed to describe the axis
orientation.  As a result, the natural description of a tangle that includes
the orientation of the axis is by \emph{two} complex numbers that depend on
space and time
\begin{equation}
	\Psi(x,t) = \sqrt{\rho}{\rm e}^{\alpha/2} \csmatl{c} \cos
	(\beta/2) {\rm e}^{i \gamma/2} \\ i \sin (\beta/2) {\rm e}^{-i
	\gamma/2} \csmatr \cv
\end{equation}
which is the natural description of the non-relativistic wave function.  Due
to the belt trick, the expression only contains half angles.  And due to the
half angles, the two-component wave function is a \iin{spinor} \cite{cs3}.
This is the strand basis of the two-component wave function that is used in
the Pauli equation since over half a century \cite{bstiomno}.

In the relativistic case, four additional functions of space are needed to
describe the wave function.  One additional function, a phase, describes with
what probability the time-averaged local belt-trick is performed left-handedly
or right-handedly.  Three additional functions describe the relativistic boost
parameters (or, if preferred, three additional Euler angles).  In total, this
doubles the functions used in the Pauli equation.

In total, the strand model requires 8 real functions of space to describe the
local wave function of spin 1/2 particles, of which 1 behaves like a density
and 7 behave like phases.  These 8 real functions can be organized as 4
complex functions of space and form what is usually called a Dirac spinor
\cite{baylistt}.  

In summary, the strand model describes wave functions as time-averaged
oriented crossing densities.  Or shorter: \emph{wave functions are blurred
tangles}.

%-----------------------------------------------------------------------------
% Feb 2009
\subsection{Dirac's equation}

Already a long time ago \cite{bpr} it was shown that the \emph{belt trick
implies the Dirac equation}.  Battey-Pratt and Racey deduced this result by
exploring a rotating object connected by strands (tails) to spatial infinity,
in the way shown in \figureref{i-belttrick}.  In their approach, the rotating
object plus the (unobservable) tails would correspond to a microscopic
particle.  (In the strand model, the central object becomes the tangle core.)
An object that is continuously rotating is described by a phase; Battey-Pratt
and Racey could show that this phase obeys the Dirac equation for free
particles, if antiparticles are added.

A simple way to see the equivalence, though different from the argument by
Battey-Pratt and Racey, is the following.  We imagine that the tails are not
observable, that the central object is negligibly small, and that it defines
the position of the microscopic particle.  In this case, a continuous rotation
of the central object corresponds to Feynman's rotating little arrow in his
well-known popular book on QED \cite{feynqed}.  Because of the tails, the
central object obeys spinor statistics and spinor rotation behaviour, as we
saw above.  Thus the tails, despite being unobservable, lead to the typical
interference behaviour of a spin 1/2 particle.  In other words, the path
integral description of quantum theory follows directly from Battey-Pratt and
Racey's approach.

In the strand model, the central object is simply the tangle core, and it is
assumed that its effective size, when tightened, is of the order of a few
Planck lengths, which makes elementary particles point-like for all practical
purposes.  Using tangles and crossing switches for the derivation of the Dirac
equation also has the advantage to introduce $\hbar$ in a natural way.

In other words, the Dirac equation results from fluctuating tangles.  The
Dirac equation describes how time-averaged fluctuating tangles evolve over
time.   In particular,
antiparticles are tangles rotating in the opposite direction, and $C$, $P$ and
$T$ transformations can be modelled in terms of strands \cite{bpr, cs3}.

In other terms, both Battey-Pratt and Racey and the strand model state that
the free Dirac equation arises naturally when looking for a description of the
belt trick that is valid for infinitesimally small angles.  In the past, the
relation between the belt trick and the free Dirac equation has been a
curiosity without physical consequences.  However, we now argue that the belt
trick for \emph{interacting} tangles can be used to deduce the three known
gauge groups.  To see this connection, we first recall the relation between
strands and Lagrangians.

%-----------------------------------------------------------------------------
% Feb 2009
\subsection{Lagrangians and the principle of least action}

What we call \textbf{action} in physics is, in the strand model, the
\emph{number of crossing switches}.  Action is thus measured in multiples of
$\hbar/2$.

In the strand model, all observed motion is due to one or several crossing
switches -- which themselves are due to change of tangle shape, induced by
strand fluctuations.  In the strand model, {quantum states} are
time-averaged \emph{tangle shapes}.  A specific average tangle shape
represents a specific quantum state.

Evolution changes quantum states, and thus tangle shape averages.  Given that
strands are fluctuating entities, and thus that all observed motion is due to
strand fluctuations, we expect that the simplest evolution, i.e., the
evolution that requires the smallest number of crossing switches, will be the
most favoured one.  The evolution with the smallest number of crossing
switches is the evolution with the smallest action.  In short, the
\textbf{least action principle} is a natural outcome of the strand model.  (In
fact, the minimization of crossing switch number leads directly to Schwinger's
quantum action principle \cite{englert}; this is an alternative way to deduce
quantum theory from the strand model.)

\textbf{Energy} is action per time.  Therefore, in the strand model,
energy is the number of crossing switches per time.  
% (In particular,
% all energy types are similar in this respect: thermal, electric,
% kinetic, magnetic, nuclear.)  
\textbf{Kinetic energy} $T$ is, in the strand model, the number of crossing
switches per time induced by shape fluctuations of tangle cores.
\textbf{Potential energy} $U$ is the number of crossing switches per time
induced by  external fields; as we will see shortly, fields are modelled by
strand loops that induce crossing switches when they approach tangle cores.

The \textbf{Lagrangian density} is the total number of crossing switches per
volume and time, averaged over many Planck scales.  This means that the
textbook definition of the free matter Lagrangian in terms of wave functions
(and their complex conjugates) and the definition of the free radiation
Lagrangians in terms of radiation fields (squared) appears naturally in the
strand model \cite{cs4}.  The Lagrangian densities of a free fermion,
\begin{equation}
    {L}_{\mathrm{Dirac}} = \bar \psi (i \hbar c \displaystyle{\not}\partial -
mc^2) \psi
\end{equation}
and of the free electromagnetic field 
\begin{equation}{L}_{\mathrm{Maxwell}} =
- \frac{1}{4\mu_0} F_{\mu \nu} F^{\mu \nu}
\end{equation}
are thus a direct consequence of
the strand model.

Special relativity is also reproduced by the strand model.  In
the strand model, the invariant limit speed $c$ in nature is given by the
(most probable) speed of a single, helically deformed, massless, i.e.,
unknotted strand.  The definition of the action 
\begin{equation}S = \int L dt\cv\end{equation}
together with the averaging procedure based on the space-time background
defined by the observer, automatically makes the action a Lorentz invariant.

In summary, it is possible to derive Dirac's equations and the Lagrangians for
free particles from the basic postulate of the strand model, and the principle
of least action is the statement that physical systems evolve with as few
crossing switches as possible.  The detailed description underlying this short
summary has been given elsewhere \cite{cs3}.

%-----------------------------------------------------------------------------
% Feb 2009
\subsection{Gauge interactions and core deformations}

The above introduction summarized how the deformation of tangle {tails} is
related to spin behaviour and Lorentz invariance.  In short, \emph{tail}
deformations induce space-time symmetries.  We will now show that, in
contrast, \emph{core} deformations induce gauge symmetries, using the
following relations:

\medskip

\begin{quotation}
\noindent \csrhd In the strand model, \textbf{gauge interactions} are modelled
by the various \emph{deformations of tangle cores} induced by external
(radiation) fields.  These deformations \emph{transfer} crossings between
boson and fermion tangles.

\noindent \csrhd In the strand model, \textbf{gauge invariance} of matter (and
radiation) fields is modelled by the invariance of tangle cores (and boson
tangles) under \emph{changes of definition} of phase.

\end{quotation}

\medskip

\noindent We will show in the following that there are \emph{three} different
ways to deform tangle \emph{cores} and to redefine core phases.
These three ways lead to exactly three gauge interactions -- and to no other.

%-----------------------------------------------------------------------------
% Feb 2009
\section{The QED Lagrangian -- U(1) gauge interaction}

\csepsf{u1}{scale=1.1111}{\emph{Twists} are core deformations whose group has a
single generator; certain tangles are affected and other are not affected by
twists, when averaged over long time scales; and random twists generate
Coulomb's inverse square law.}

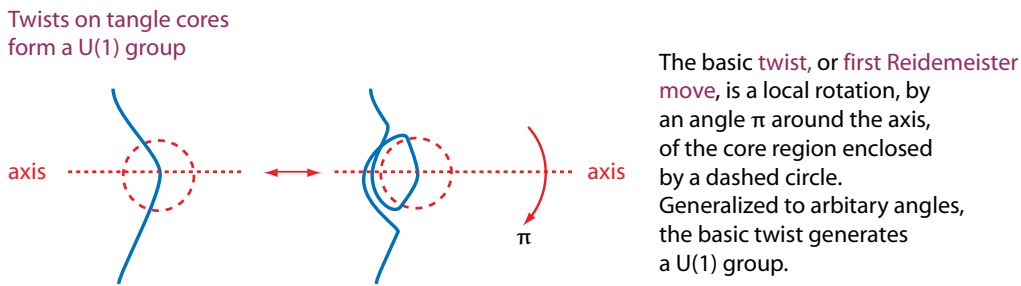
\csepsf{i-twist-circle}{scale=1}{How the set of generalized twists -- the
set of all local rotations of a single strand segment around an axis -- forms
a U(1) group group.}

In the strand model, any \emph{observable} deformation must switch
some crossings.  The simplest way to deform a tangle \emph{core}, while changing its
crossing number at the same time, is to take a piece of strand and to
\emph{twist} it.  This is shown in \figureref{u1}.  In knot theory, this type
of deformation is called the \emph{first Reidemeister move}.  We will soon
find out that the addition of a twist to a core -- or better, the \emph{transfer}
of a twist between a core and a single external strand -- has the same
properties as what is usually called the emission or absorption of a photon.

We note that a twist transfer from a loop to a core -- as shown on the top
right of \figureref{u1} -- does \emph{not} imply any cutting of strands.  An
approaching twisted loop will influence the fluctuations of the core, due to
strand impenetrability, and will lead to an effective transfer of twist,
without any `ugly' change of topology.

A twist has the same effect on a tangle core as a \emph{local rotation} by an
angle $\pi$.  The set of all local rotations, for any angle, forms a U(1)
group.  We thus conjecture that twists represent the electromagnetic
interaction.

We note that large numbers of random twists will affect certain tangle cores
while they will not affect others, as shown in the bottom left of
\figureref{u1}.  For example, if we take a single, knotted strand, we expect
that its long-time average motion will be affected by random twists only if
the knot is chiral.  In other words, knotted \emph{chiral} strands represent
electrically charged bosons, whereas knotted \emph{achiral} strands represent
neutral bosons.

In summary, sensitivity to twists suggests to define \textbf{electric charge}
with the \emph{chirality} of a particle tangle.  For example, a granny knot
represents one elementary charge.  Right and left chirality then correspond to
positive and negative charge.  As a consequence of this definition, charge
values for free particles are naturally quantized.  (A separate discussion
\cite{cs5} shows how the definition of charge is extended to quarks and to
doubly charged hadrons.)

Because all interactions are only strand deformations -- in particular, strand
interpenetration being forbidden -- total tangle chirality is conserved in all
interactions.  In short, the strand model predicts that \emph{electric charge
is conserved} in all interactions, as is indeed observed.

The strand model states that only tangles with localized cores can be charged;
only tangles with cores can be grabbed in such a way as to be twisted.  In the
strand model, knotted cores imply non-vanishing mass.  In other words, only
\emph{massive} particles can be charged, exactly as is observed.
In short, any \emph{charged} fermion is a \emph{chiral} tangle that is made of
two or more strands.

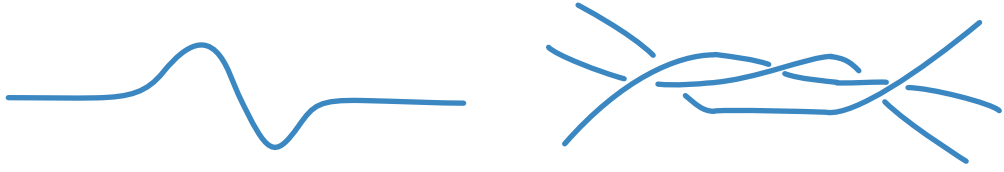
\csepsf{i-ee-fermion-photon}{scale=1}{A tangle candidate for a photon (left) 
and for a chiral lepton (right).}

Photons are helical strands.  Photons have no knotted core: they are massless
and neutral.  Photons come in two versions: with positive and negative
helicity.  Virtual photons correspond to the two types of twists that can be
added to tangle cores \cite{cs4}.  Photons cannot disappear in the strand
model.  The energy and angular momentum content of the helix is conserved.  If
one strand loses its helix, a neighbouring strand will gain it.  In the strand
model, photons are thus particles with infinite lifetime.

The particle tangle of the photon and a tangle of a charged fermion are shown
in \figureref{i-ee-fermion-photon}. We note that despite their simple 
structure, photons cannot disappear. This becomes clear once the vacuum 
strands around the photon are drawn \cite{cs4}.

To explore the translational motion of a fermion tangle, we remember that
vacuum itself is also made of strands.  To move through space, a fermion must
move through the vacuum web of strands.  Fermion tangles move through space
either by exchanging strands with the vacuum, or by `passing around' vacuum
strands at spatial infinity.  In short, the vacuum web \emph{hinders} fermion
motion; in contrast, photon motion is \emph{not} hindered by the vacuum web.
In short, charged (and massive) tangles are predicted to move more slowly than
light, as is observed.  In the strand model, the speed of light thus cannot be
attained by fermion tangles: it is a limit speed.

In the strand model, the physical action of a process measures the number of
required crossing switches; each crossing switch makes a contribution of
$\hbar/2$.  For a charged fermion in an electromagnetic field, the action will
be given by three terms: the crossing switches due to the motion of the
fermion, those due to the motion of the electromagnetic field, and those due
to the interaction.
% This relation defines the sign of the three terms in the
% Lagrangian of QED. 

In the strand model, like in quantum field theory, the  form of the
electromagnetic interaction term is fixed by gauge invariance.

%-----------------------------------------------------------------------------
% \subsubsubsubsubsubsubsubsection{Table gauge-shape correspondence}
\begin{table}[p]
\small
\caption{The correspondence between shape change of tangles and gauge theory}  
\protect\label{catgaugetab}
\centering
\begin{tabular*}{\textwidth}{@{} >{\PBS\raggedright} p{30mm}
@{\extracolsep{\fill}} >{\PBS\raggedright} p{50mm} 
@{\extracolsep{\fill}} >{\PBS\raggedright} p{50mm}  @{}}
\hline
\textsc{Concept}  &  \textsc{Gauge theory} & \textsc{Strand model}\\[0.5mm]
\hline
\textbf{System} & Matter \& gauge field  & Tangle core \& approaching 
loop \\
\textbf{Interaction} & Phase change by field &  Core deformation by loop  \\
\textbf{Gauge freedom} & Freedom to define {quantum phase} and vector
potential & Freedom to define core phase and boson energy-momentum
density \\
Gauge transformation & Changes definition of quantum phase and vector potential
& Changes phase value of
core and loop energy-momentum
density  \\
Gauge-dependent  & Quantum phase,
& Core's angular orientation,\\
quantities & vector potential, &  loop
energy-momentum
density, \\
& change of vector potential along \emph{open} path
	      & orientation change along an \emph{open} path in (core) deformation
	      space  \\
\textbf{Gauge-independent} %
           & Field strength,  & Crossing density and flow,  \\
\textbf{quantities} 	   & phase difference along a \emph{closed} path,
& core phase difference after \emph{closed} path
in deformation space, \\
& integral of vector potential along a \emph{closed} path & loop
energy-momentum density along \emph{closed} path\\
Gauge boson & Gauge group generator & Deformation generator \\  
Gauge group origin & Unknown
& Due to possible core deformations, 
classified by Reidemeister moves \\
Gauge groups     & U(1)        & \emph{Twisting} a 
core strand \\
     & SU(2)        & \emph{Poking} a core strand under another \\
     & SU(3)        & \emph{Sliding} a strand across two others \\
Charge  &  Couples matter to 
field: & Topological property that yields:\\  
& -- electric charge & -- preferred sensitivity to one sign of twist: chirality  \\
& -- weak charge & -- preferred sensitivity to one sign of poke: tangledness
with helicity \\
& -- colour charge  & -- preferred sensitivity to one sign of slide: rationality \\
Coupling strength  & Coupling constant & Probability of core deformation\\
\hline
\end{tabular*}
\vss
\end{table}

%-----------------------------------------------------------------------------
% Feb 2009
\subsection{U(1) gauge invariance}

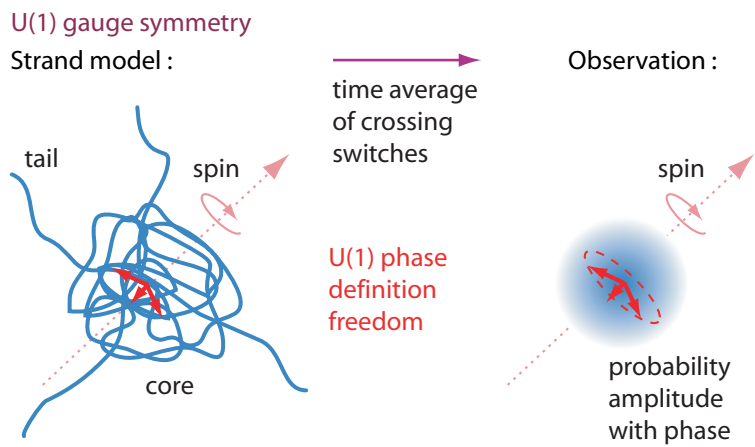
\csepsf{i-phasedef}{scale=1}{Various ways of defining a tangle
orientation, showing the explicit U(1) gauge freedom for one such
definition.}

In 1984, Berry, Wilczek, Zee and Shapere deduced a well-known result about the
motion of deformable bodies: the freedom to define a measure of deformation
leads to a gauge theory \cite{berry, shapwil, squarecat}.  The main points of
this equivalence are summarized in \tableref{catgaugetab}.  In particular, the
studies showed that the freedom of defining a measure of deformation is
analogous to the freedom of choosing a gauge.  As in gauge theory, also in the
study of body deformations there are gauge-dependent and gauge-independent
quantities.  In particular, if a sequence of deformations returns a system
back to its original state, this process allows to define a quantity that is
\emph{independent} of the chosen deformation measure, and thus
gauge-independent.

In the strand model, \textbf{interactions} are transfers of observable
deformations, i.e., of crossing switches, between boson and fermion cores.  In
particular, twists \emph{change the phase} of the fermion core, as shown in
\figureref{u1}.  Twists can be concatenated; and they have a single generator.
When concatenated twists reach a total core rotation by $4 \pi$, the system
can fluctuate back to its initial state.  Twists thus generate a U(1) group.
The corresponding gauge-invariant quantity is the number of turns of the core,
in short, the \emph{phase difference}.

On the other hand, the \emph{absolute phase} of a tangle core is not uniquely
defined.  For example, the absolute phase could be defined by the direction
formed by the outermost crossing with respect to the core centre.  But any
other direction along a circle perpendicular to the rotation axis is also
possible, as shown in \figureref{i-phasedef}.  The freedom in the definition
of the phase again corresponds, once the axis is given, to a U(1) group.  

The other important gauge-independent quantities in this system are the volume
density and the flow density of twists; this is precisely the electromagnetic
field intensity, and that Maxwell's equations follow from the twist density
and flow in the macroscopic limit \cite{cs4}.  In particular, Coulomb's law is
a consequence of the random emission, by charged fermions, of twisted loops
into all directions of space.  This is illustrated in \figureref{u1}.  In
particular, particles of one handedness prefer to emit virtual photons of one
handedness.  As a result, particles of the same handedness repel, and
particles of different handedness attract.

We thus recover several well-known properties of charged quantum particles:
(1) electromagnetic fields change particle phase, (2) only phase differences,
but not absolute phase values, can be measured, (3) charged particles move
slower than light and attractor repel, and (4) there is a U(1) gauge
invariance for transfers of twists, i.e., minimal coupling holds.

%-----------------------------------------------------------------------------
% Sep 2006, Feb 2009
\subsection{Quantum electrodynamics}

\figureref{u1} shows the transfer of twists between bosons and fermions that
results from the mutual hindrance of strand fluctuations.  The figure also
shows that random (virtual) twist emission leads, after averaging over space
and time, to Coulomb's law.  Charge conservation, Coulomb's law and its
generalization for relativistic observers leads to Maxwell's equations.  In
short, by averaging twist numbers over space and time we obtain the Lagrangian
of the free electromagnetic field \cite{cs4} 
\begin{equation}{L}_{\mathrm{Maxwell}} = - \frac{1}{4\mu_0}
F_{\mu \nu} F^{\mu \nu} \cp\end{equation}
%
% As explained above, free fermion tangles are described by the Dirac
% Lagrangian.  
In addition, the U(1) gauge invariance of twist exchange between charged
fermions adds an interaction term to the free Dirac Lagrangian found above.
We thus get the complete Lagrangian of QED from strands
\begin{equation}{L}_{\mathrm{QED}} = \bar \psi (i \hbar c \displaystyle{\not}
D - mc^2) \psi - \frac{1}{4\mu_0} F_{\mu \nu} F^{\mu \nu}\cv\end{equation}
where $\displaystyle{\not} D$ is the U(1) gauge-covariant derivative
describing the electromagnetic interaction.

Another way to see this result is the following.  Our discussion showed that
if we identify core twists with the electromagnetic interaction, we can deduce
the following statements:

\medskip

-- electric charge is conserved and quantized;

-- electric charges move slower than light;

-- all motion is described by the quantum of action $\hbar$;

-- photons are massless;

-- charges emit and absorb virtual photons;

-- electromagnetic fields change the phase of charged particles;

-- the QED Lagrangian has a U(1) gauge symmetry.

\medskip

\noindent When these statements are added to the free field Lagrangian, we get
the full Lagrangian of QED. 

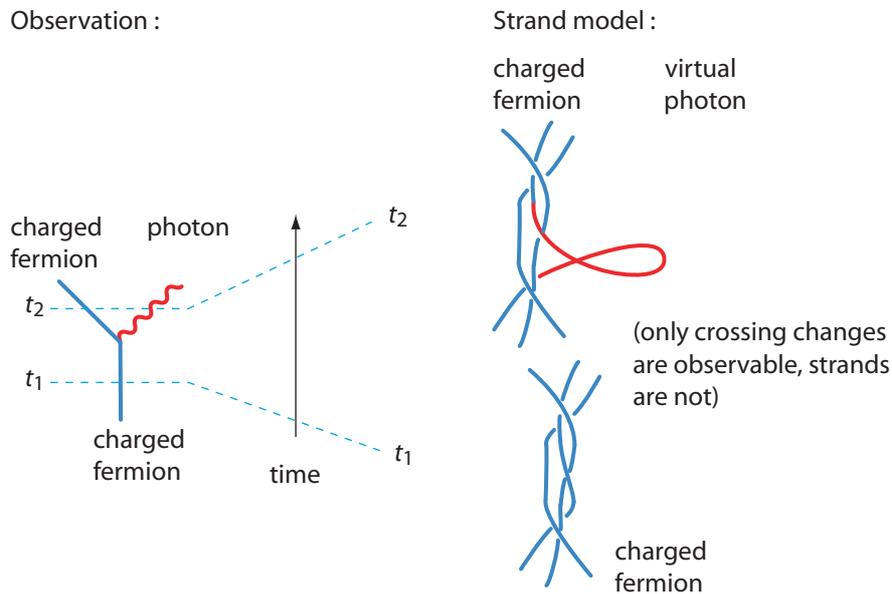
\csepsf{i-ee-qed-vertex}{scale=1}{The fundamental Feynman diagram for
QED for a specific time orientation (left) and one of the corresponding
strand diagrams (right), with different magnification scales.}

% \csepsf{i-qeddetails}{scale=0.55}{The various cases of the fundamental
% Feynman diagram of QED and their strand counterparts.}

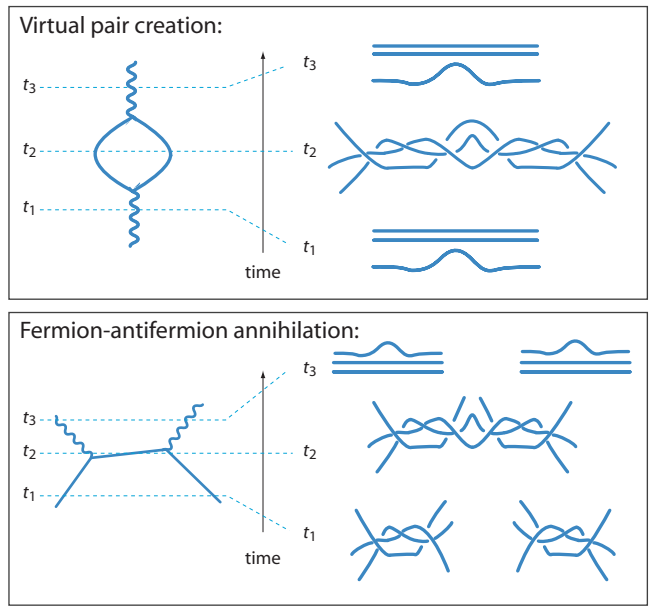
\csepsf{i-allqed}{scale=0.5}{A few QED processes described in terms of 
strands.}

We can check the equivalence between QED and the strand model in many ways.
It is well-known that all of quantum electrodynamics follows from its
fundamental Feynman diagram, shown on the left of \figureref{i-ee-qed-vertex}.
The strand model provides a corresponding diagram, shown on the right of the
figure.  (For clarity, the magnification scales are chosen differently.)  
% A
% full exploration of the fundamental Feynman diagram is shown in
% \figureref{i-qeddetails}.  We note that sometimes a deformation involving
% \emph{spatial infinity} is necessary to describe the process.  
On this basis, strand diagrams for all known QED Feynman diagrams can be
constructed; two examples are shown in \figureref{i-allqed}.  

We can check the equivalence of strand deformations and Feynman graphs also
through their conservation properties.  A twist conserves tangle topology.
Therefore every twist, i.e., every electromagnetic reaction, is predicted to
conserve the spin of the involved particle, and the total spin of the system.
Since unchanging topology also implies unchanging \textbf{flavour} (as we will
see below) the strand model predicts that the electromagnetic interaction
conserves particle flavour.  Both properties agree with observation.

Another quantity of interest is \textbf{intrinsic parity} $P$.  Parity is
related to the behaviour of tangles under reflection.  In the strand model,
parity $P$ is a topological quantity that distinguishes a fermion from an
antifermion.  The strand model automatically implies that fermions and
antifermions have \emph{opposite} intrinsic parity, as is observed.  The
strand model also predicts that photons 
% !!!2 bosons?
and their `antiparticles' have the
\emph{same} intrinsic parity, as they can be deformed into each other.  This
is indeed observed.  Since twists conserve topology, the strand model predicts
that parity is conserved in electromagnetic reactions, as is indeed observed.

% SIDE POINT: The strand model
% predicts that the skein for the photon has negative parity (as reflection
% reverses the orientation of the helical photon screw), as observed.

\textbf{Charge conjugation} parity or $C$-parity is the behaviour of tangles
under charge conjugation.  In the strand model, charge conjugation is the
exchange of all crossings \cite{cs3}.  The strand model thus implies that only
neutral particles can have a defined $C$-parity value, as is observed.  The
strand model also predicts that the photon tangle has negative $C$-parity, as
observed.  Finally, the strand model predicts that the electromagnetic
interaction conserves $C$-parity for the same reason that it conserves
$P$-parity.  This is also observed.

%-----------------------------------------------------------------------------
% Feb 2009 
\subsection{Renormalization of QED}

In quantum field theory, Lagrangians must not only be Lorentz and gauge
invariant, but must also be renormalizable.  The strand model makes several
statements on this issue.  At this point, we focus on QED only; the other
gauge interactions will be treated below.  First fo all, the strand model
reproduces the observation that only one basic Feynman diagram exists for QED.
In other words, the strand model of QED is equivalent to usual QED, and thus
is renormalizable.  The strand model thus only provides a new underlying
picture for Feynman diagrams; the strand model does not change the physical
results at any experimentally accessible energy scale.  In particular, the
measured running of the fine structure constant and of the masses of charged
particles are reproduced by the strand model, because Feynman diagrams of all
orders are reproduced.

The twist deformations underlying the strand model for QED suggest new
ways to calculate effects of higher order Feynman diagrams, such as needed
in calculations of g-factors of charged particles.  In particular, the strand
model for QED, as shown in \figureref{i-ee-qed-vertex}, suggests that higher
order QED diagrams are simple deformations of lower order diagrams.  Taking
statistical averages of strand deformations thus in principle allows to
calculate QED effects to arbitrary order in the coupling.  However, this topic
is not part of the present paper.

The strand model also suggests that the difference between renormalized and
unrenormalized quantities reflects the difference between minimal and
non-minimal crossing switch numbers, or equivalently, between simple and more
complex, small-size tangle deformations.  In more detail, unrenormalized
quantities can be imagined as those deduced when the tangles are pulled tight,
whereas renormalized quantities are those deduced for particles surrounded by
many large-size fluctuations.

%-----------------------------------------------------------------------------
% Sep 2006, Feb 2009
\subsection{Predicted limit values and deviations from QED}

The equivalence of QED Feynman diagrams and strand diagrams implies that
deviations of the strand model from QED are expected \emph{only} when
short-time fluctuation averaging is not applicable any more; this happens only
when quantum gravity starts to play a role.  This will only happen near the
Planck energy $\sqrt{\hbar c^5/4G}$.

The strand model predicts that all Planck units are \emph{limit} values.  For
example, in the same way that the maximum speed is $c$, also the maximum
elementary particle energy is the Planck energy and the shortest measurable
length is the Planck length $\sqrt{4G\hbar/c^3}$.  This view yields a maximum
electric field value $E_{\rm max} = {c^4}/{4Ge}\approx {2.4\cdot10^{61}}\,{\rm
V/m}$ and a maximum magnetic field value $B_{\rm max} = { c^3}/{4Ge} \approx 8
\cdot 10^{52}\,\rm T$ \cite{cs4}.  All physical systems -- including all
astrophysical objects such a gamma ray bursters or quasars -- are predicted to
conform to this limit.  These limit values form another way to characterize
the domain where deviations of the strand model from ordinary QED are
expected.

All limit values for observations have a simple explanation: limit values
appear when strands are as closely packed as possible.  In the strand model,
strands cannot be packed more closely than to Planck distances.

In summary, the strand model suggests that U(1) invariance is valid for all
energies below the Planck energy; no other gauge group at higher energy is
predicted to appear, and grand unification is ruled out.

%-----------------------------------------------------------------------------
% Sep 2006, Feb 2009
\section{The weak interaction Lagrangian -- broken SU(2) gauge invariance}

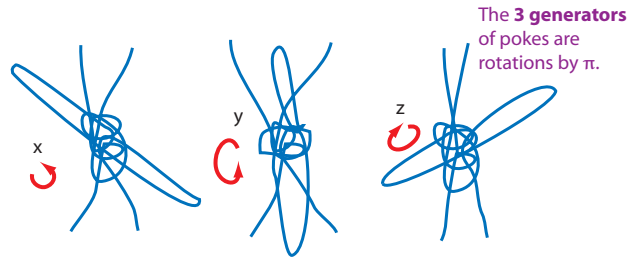
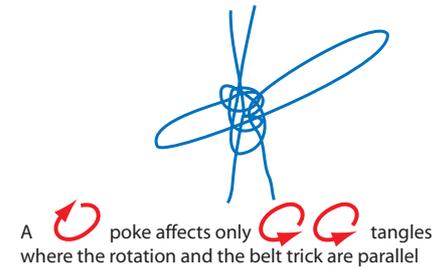
\csepsf{su2}{scale=0.75}{\emph{Pokes} as core deformations, as
elements of a SU(2) group, and some tangles that are affected and some
that are not affected by pokes.}

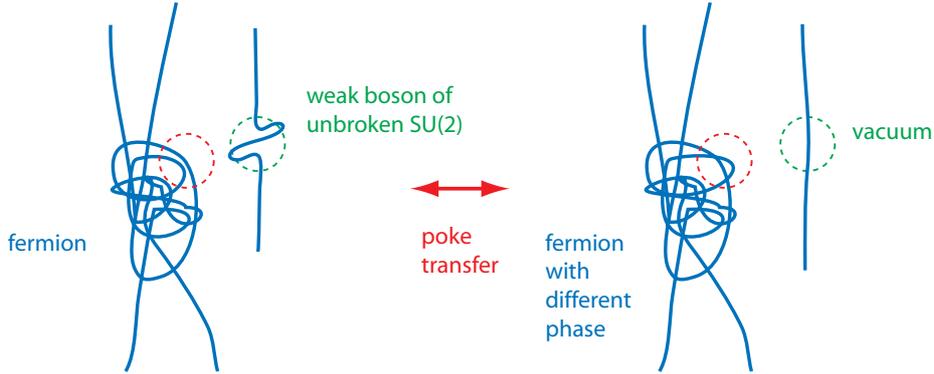
\csepsf{i-strands-weak}{scale=1}{\emph{Poke transfer} is the basis of the
weak interaction in the strand model.  No strand is cut or reglued; the
transfer occurs only through the excluded volume due to the impenetrability of
strands.}

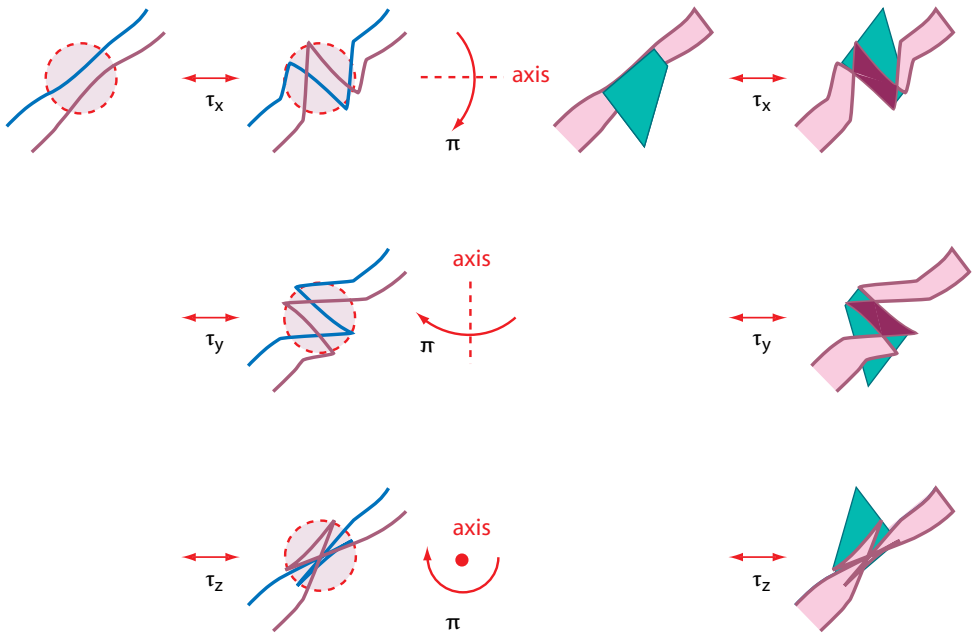
\csepsf{i-poke-cube}{scale=1}{How the set of all pokes -- the set of all
deformations induced on tangle cores by the weak interaction -- forms an SU(2)
gauge group.  The relation to the belt trick, with a pointed buckle and two
belts, is also shown.}

The next simplest way to deform a tangle core in such a way that the
crossing number changes is to \emph{poke} a piece of strand over a
second piece of strand, as shown in \figureref{i-strands-weak}.  In knot theory,
this type of deformation is called the \emph{second Reidemeister
move}.  We will find out that the poke deformation of a core
represents what is usually called the emission or absorption of a weak vector boson.

A poke has the same effect as a localized, partial rotation (plus a possible
size extension) of the tangle core.  The simplest way to see this is to
imagine a large number of pokes acting all over a tangle core: the core will
be rotated (and possibly extended).  Below, we will explore the exact
conditions for this to happen.

Pokes can be concatenated and form a group.  In particular, the three pokes
around the three orthogonal axes, shown in \figureref{i-poke-cube},
do not commute.  Closer inspection shows that the three pokes are equivalent
to the operations involved in the belt trick; the commutator of two orthogonal
pokes is the third poke (multiplied by $-1$ or $+1$ depending on the
permutation).  In addition, performing the same poke twice, we get a twisted
situation; in the strand model this is represented by a $-1$ \cite{cs4}.  The
generators thus obey
$$
 \let\l\tau 
\begin{array}{c|ccc} 
  \cdot & \l_x    &    \l_y &    \l_z  \\  
\hline % instead of midrule 
\l_x    &   -1    &  i\l_z  & -i\l_y   \\ 
\l_y    & -i\l_z  &  -1     &  i\l_x   \\ 
\l_z    &  i\l_y  & -i\l_x  &   -1     \\ 
\end{array}
$$
and the
group formed by pokes is thus SU(2).
% More precisely, any tangle transforms under pokes as a fundamental and
% faithful representation of SU(2), and its Lagrangian is SU(2) invariant.  
We thus conjecture that deformations by pokes represent the weak interaction.

As a note, it is worth recalling that the (broken, as we will see) SU(2) gauge
symmetry of the weak interaction is realized in the strand model by the
deformation of the tangle \emph{cores}, whereas the (unbroken) SU(2) group due
to the Pauli matrices due to spin 1/2 is realized through deformations of the
tangle \emph{tails}, keeping the core rigid.  The two structures are
independent.

The properties of pokes and their SU(2) gauge group differ from the U(1)
twists in four aspects: they can change topology, they violate parity, they
interact among themselves and they break the SU(2) symmetry.

%-----------------------------------------------------------------------------
% Feb 2009
\subsection{Particle transmutation and topology change}

If a poke that is applied to a fermion tangle involves spatial infinity (thus
if a loop goes `over' a tail at spatial infinity) the move can change the
topology of the fermion.  In the strand model, a different tangle topology
means a different particle.  There is nothing preventing this process at
spatial infinity, as background space is not defined there.  In short, the
strand model predicts that weak interactions can transform different particles
into each other.  This is observed, e.g., in beta decay.  In contrast, the
electromagnetic twists discussed above never have this effect.

We note that the properties of pokes imply that despite particle
transmutation, they conserve total electric charge, spin, and weak charge, as
is observed.

%-----------------------------------------------------------------------------
% Feb 2009
\subsection{Parity violation and weak charge}

Certain tangle cores will be affected by large numbers of similar pokes,
whereas others will not be.  For example, if we take a \emph{single}, knotted
strand, we see that it will be rotated.  In other words, single knotted
strands, or massive bosons, are weakly charged.  We will find out shortly that
this means that the W and the Z, in contrast to photons, interact among
themselves.

Let us explore a fermion tangle, in this case a rotating tangle core made of
\emph{two} strands -- as shown in the bottom right of \figureref{su2} -- that
is subject to a large number of similar pokes.  We first recall that for a
given rotation by $4 \pi$, the belt trick can be performed in \emph{two} ways,
parallel and antiparallel to the sense of rotation.  (An animation showing the
two options in a clear way is available on the internet \cite{egan}.)  On
average, a poke will act in one way on states where rotation and belt trick
sense are \emph{parallel}, and will act in another way -- namely not at all --
on states where rotation and belt trick sense are \emph{antiparallel}.  We
thus find that fermions are affected by pokes depending on their their
helicity.  Antiparticles, which are represented by tangles rotating backwards,
will only be affected by those pokes which do not affect particles.  In short,
only tangles of one handedness are weakly charged; tangles of the other
handedness have no weak charge.  Pokes thus reproduce the observed maximal
parity violation of the weak interaction.

% Weak interactions are not parity invariant.  Experimental observations can be
% summarized in the following way \cite{weisskopf}: the \emph{only} helicity
% states of extremely relativistic fermions that participate in charge-changing
% weak interaction processes are $-1/2$ for particles and $1/2$ for
% antiparticles.  In other words, parity is observed to be violated maximally by
% charged current weak interactions.

We note that weak charge -- the weak isospin -- requires localization of a
tangle.  In other words, the strand model predicts that only massive fermions
can interact weakly, as is observed.

We also note that the ability to interact weakly does not depend on the
detailed tangle topology, but only on tangledness.  All fermions of a given
handedness have the same weak charge (i.e., the same value for the third
component of the weak isospin).  The strand model thus predicts that all
left-handed matter fermions (respectively, all right-handed antifermions) have
the \emph{same} value of the weak isospin (respectively, the same negative
value), as is observed.

$C$-parity violation by the weak interaction pokes appears in the same way as
$P$-parity violation.  In contrast, electromagnetic twists do not violate
either $P$ or $C$-parity.  (The observation of a small $CP$ violation by the
weak interaction is discussed elsewhere \cite{cs5}.)

% \end{document}
% \endinput

%-----------------------------------------------------------------------------
% Feb 2009
\subsection{Massive gauge bosons}

% \csepsf{i-ee-weak-intvec-bosons}{scale=0.9}{The strand models for the
% neutral weak intermediate vector boson $\rm Z^{0}$ (left, measured to
% have a mass of 91.2\,GeV) and its charged counterpart $\rm W$ (right,
% measured to have a mass of 80.4\,GeV)}

The third difference between SU(2) pokes and U(1) twists concerns the
associated gauge bosons.  If we apply pokes that involve spatial infinity to
the high-energy boson tangle, we get two candidates for the low energy tangles
of the W and Z bosons, shown in \figureref{i-strands-su2breaking}.  Their tangles
consist of a single knotted strand: they are thus massive.  The chiral trefoil
tangle represents the charged W boson, its mirror version the corresponding
antiparticle of opposite charge; the achiral figure-eight tangle represents
the Z boson.  (More complex tangles of one strand represent states of even
shorter lifetime with added virtual W or Z particles.)

The mass of the weak vector bosons is an essential property; it explains the
weakness of the weak interactions.  Since the tangles for the W and the Z are
different, the strand model also reproduces their difference in mass.

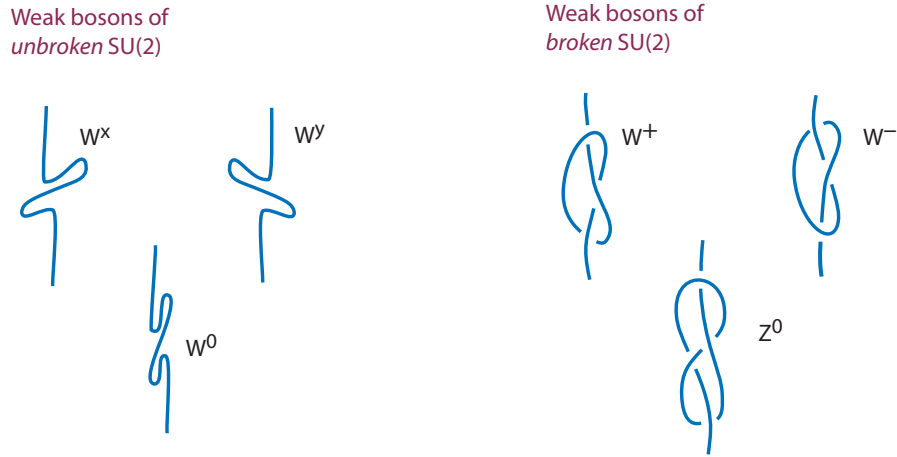
\csepsf{i-strands-su2breaking}{scale=1}{Virtual, high-energy, unbroken
SU(2) boson tangles (left) and real, low-energy W and Z boson tangles
that break SU(2) (right). Poke-inducing strands (left) differ
from weak vector bosons (right) because of symmetry breaking.  The figure
shows only the simplest possible tangles for each weak gauge boson.}

The trefoil tangle of the $W$ has spin 1; it is chiral, thus is electrically
charged.  It is knotted, thus has non-zero mass.  The tangle has no $P$ and
$C$ parity, as is observed.  The figure-eight tangle of the $\rm Z$ has spin
1; it is not chiral, thus is electrically neutral.  It is knotted, thus has
non-zero mass.  The tangle is its own antiparticle, as is observed.  The
tangledness of the W and Z tangles also implies that they couple to pokes,
thus to themselves.  The two tangles thus have non-vanishing weak isospin and
lead to a non-Abelian gauge theory, a Yang-Mills theory, for the exchange of
pokes.

%-----------------------------------------------------------------------------
% Feb 2009
\subsection{SU(2) breaking and mass generation}

The $\rm Z$, the $\rm W^+$ and the $\rm W^-$ bosons can be seen as a broken weak
isospin triplet representation of the SU(2) gauge group of the weak
interaction.  The degeneracy is explicitly broken in the strand model: the
differences in shape -- e.g., in crossing numbers -- of the two tangles are the reason
for the symmetry breaking.

\figureref{su2} illustrates the weak interaction with the help of
small-amplitude poke moves.  In other words, the moves shown in that figure
correspond to fluctuations at very high energy.  At such energies, the SU(2)
gauge symmetry is predicted to be exact, or unbroken, by the strand model.
(This explicitly contradicts grand unified theories, which predict higher
symmetry groups at high energy.)

At low energy, the pokes show effects due to the involvement of spatial
infinity: at low energy, topology changes due to such poke moves thus play a
role.  This difference between high and low energy, the breaking of SU(2)
symmetry, is illustrated in \figureref{i-strands-su2breaking}.  
% The possibility to
% apply pokes to high-energy bosons tangles leads to knotted low-energy boson
% tangles.  
Knotted tangles are massive, and different knots have different masses.  Thus
the strand model predicts massive bosons and the breaking of SU(2) symmetry at
low energy.

In other terms, the strand model predicts that a W or a Z boson is described
by a large number of tangles: a tangle without a knot, a tangle with a simple
knot, and an infinite number of tangles with more complex knots.  In addition,
the strand model predicts that tangles with (on average) more complex knots
are more massive that tangles with simpler knots.  This explains why the
neutral Z boson is heavier than the charged W boson.

In summary, the strand model predicts that mass is a result of tangledness,
and that mass generation (for bosons and for fermions) is related to the weak
interaction.

% %-----------------------------------------------------------------------------
% % Sep 2006
% \subsection{Neutrinos}
% 
% % \csepsf{i-ee-neutrino-electron}{scale=0.7}{Candidate tangles for the neutrino
% % (left) and electron (right)}
%  
% One of the recent results of particle physics is that neutrinos have
% non-vanishing mass.  The strand model makes the same suggestion; first, the
% model states that all elementary fermions are localized tangles of two or
% three strands \cite{cs5}, and no such structure can be massless.  Secondly,
% the strand model requires all weakly interacting fermions to have mass, as
% explained above.  The strand model thus provides two arguments for neutrino
% masses.
% 
% The tangle structures of neutrinos will be discussed in
% more detail in the forthcoming paper \cite{cs5}, which provides the tangle
% models for all elementary particles.  It suffices here to mention that a
% suitable poke move involving infinity can transform the candidate neutrino
% tangle to the candidate electron tangle, as required by the weak interaction.
% 
% The neutrino tangle turns out to be achiral, thus with zero charge.  It is linked and
% localized, thus with non-zero mass, and it differs from the antineutrino 
% tangle. 
% The strand model thus suggests that neutrinos are Dirac particles.  (This
% topic will be expanded in the forthcoming paper \cite{cs5}.)  This prediction
% has not yet been tested and is not contradicted by experiments, which so far
% all tested high-energy neutrinos.

%-----------------------------------------------------------------------------
% Feb 2009
\subsection{QFT of the weak nuclear interaction}

% In the strand model, the physical action of any process measures the number of
% required crossing changes.  In case of the weak interaction, crossing changes
% appear through the poke process induced by virtual or real vector bosons.  
We can summarize the equivalence between pokes and the weak interaction in the
following statements:

\medskip

-- weak charge is conserved, related to handedness and is defined for tangles;

-- the weak charge is the same for all elementary fermions;

-- the weak charge is different for bosons;

-- the weak interaction violates $P$ and $C$ parity maximally;

-- the weak intermediate vector bosons have mass;

-- the weak interaction has a SU(2)  gauge symmetry that is broken at 
low energy;

-- weak charges emit and absorb virtual intermediate vector bosons.

\medskip

\noindent All these statements, together with the Dirac equation, reproduce
many terms of the electroweak Lagrangian.  In particular, the SU(2) gauge
invariance defines the interaction terms between the weak fermionic doublets
and the weak interaction bosons.

Several terms and aspects of the electroweak interaction Lagrangian are not
yet reproduced: (1) the terms involving the Higgs, (2) the quark and neutrino
mixing matrices, (3) the number of generations, (4) the particle masses.
These aspects are discussed elsewhere \cite{cs5}.

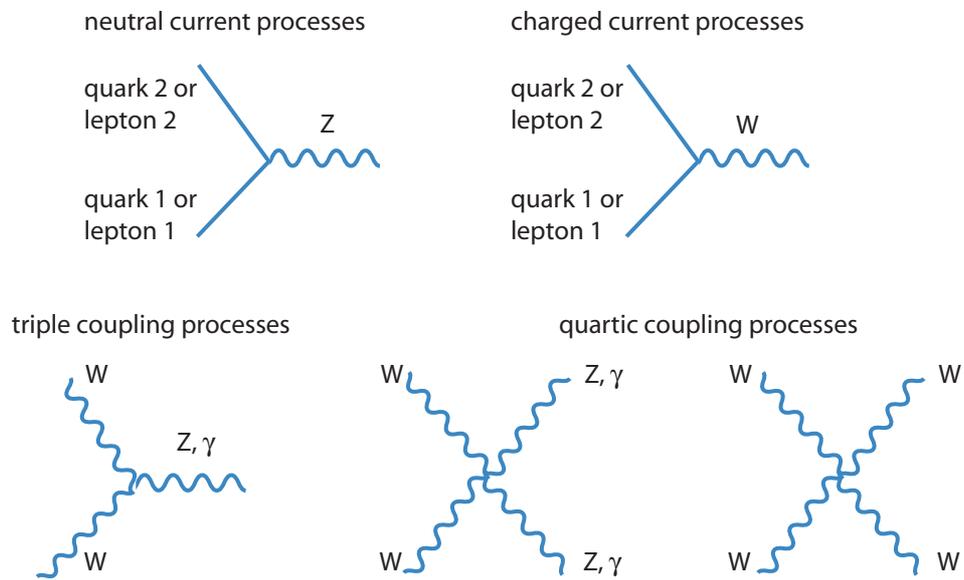
\csepsf{i-ee-weak}{scale=1}{All the Feynman diagrams for the weak interaction
that do not involve the Higgs field.}

But those parts of the electroweak Lagrangian that \emph{are} reproduced so far can be
checked further.  At energies lower than the electroweak unification scale,
reactions due to the weak interaction can be classified into neutral and
charged current processes, as well as triple and quartic boson couplings.
This is shown in \figureref{i-ee-weak}.  To describe these processes in the
strand model, we need the tangles of the real and virtual intermediate vector
bosons $\rm W$ and $\rm Z$ from \figureref{i-strands-su2breaking}.  The resulting
strand model for neutral currents is shown in \figureref{i-ee-qad-neu-vertex}.

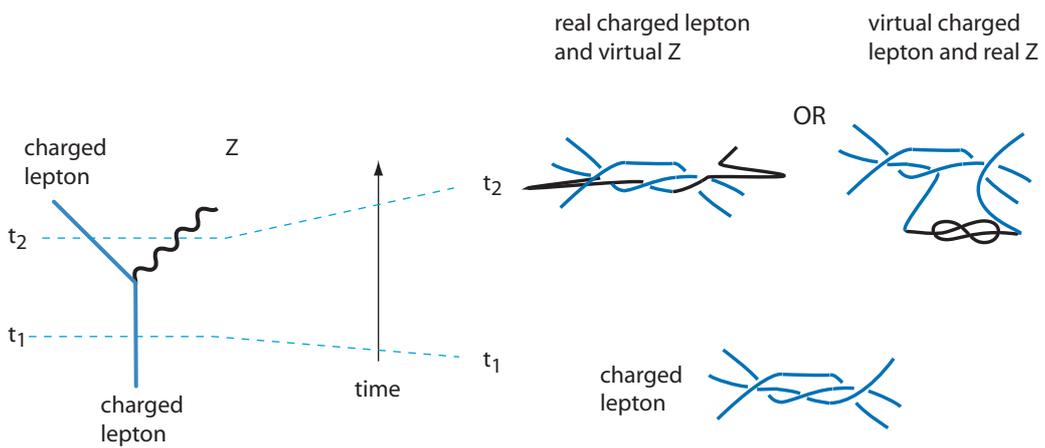
\csepsf{i-ee-qad-neu-vertex}{scale=1}{The Feynman diagram for a weak \emph{neutral}
current interaction (left) and the corresponding strand diagram (right).}

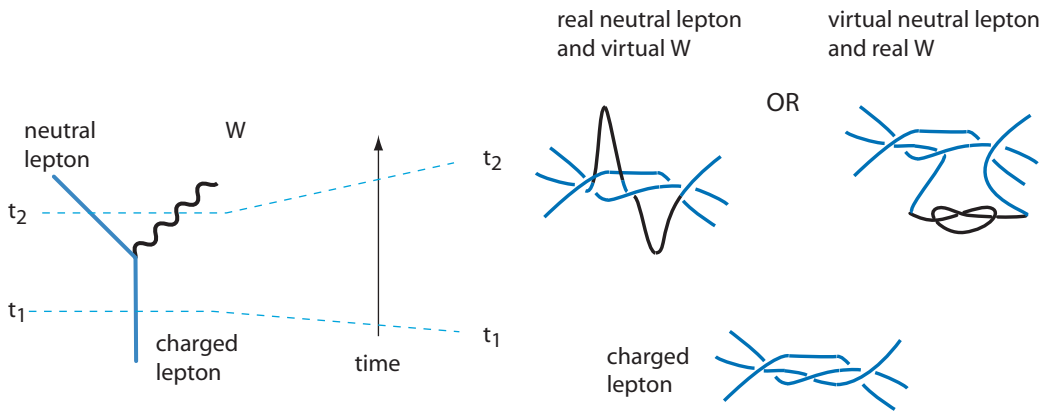
\csepsf{i-ee-qad-cha-vertex}{scale=1}{The Feynman diagram for a weak
\emph{charged} current interaction (left) and the corresponding strand diagram
(right).}

The neutral currents can be reduced to two Feynman diagrams: a leptonic vertex
and a hadronic vertex, as shown in the upper left corner of
\figureref{i-ee-weak}.  In the strand model, leptonic \emph{neutral} currents
leave the topology of the interacting matter particles unchanged.  In this
way, weak neutral currents are automatically flavour-conserving in the strand
model, as is observed.

\figureref{i-ee-qad-cha-vertex} for the \emph{charged} currents requires a
move that involves spatial infinity and then changes the topology of the
involved fermions.  However, the process only changes leptons into leptons and
quarks into quarks, as is observed.  (Hadronic neutral and charged currents,
and tangles for all leptons, quarks and the Higgs boson, are discussed 
elsewhere \cite{cs5}.)

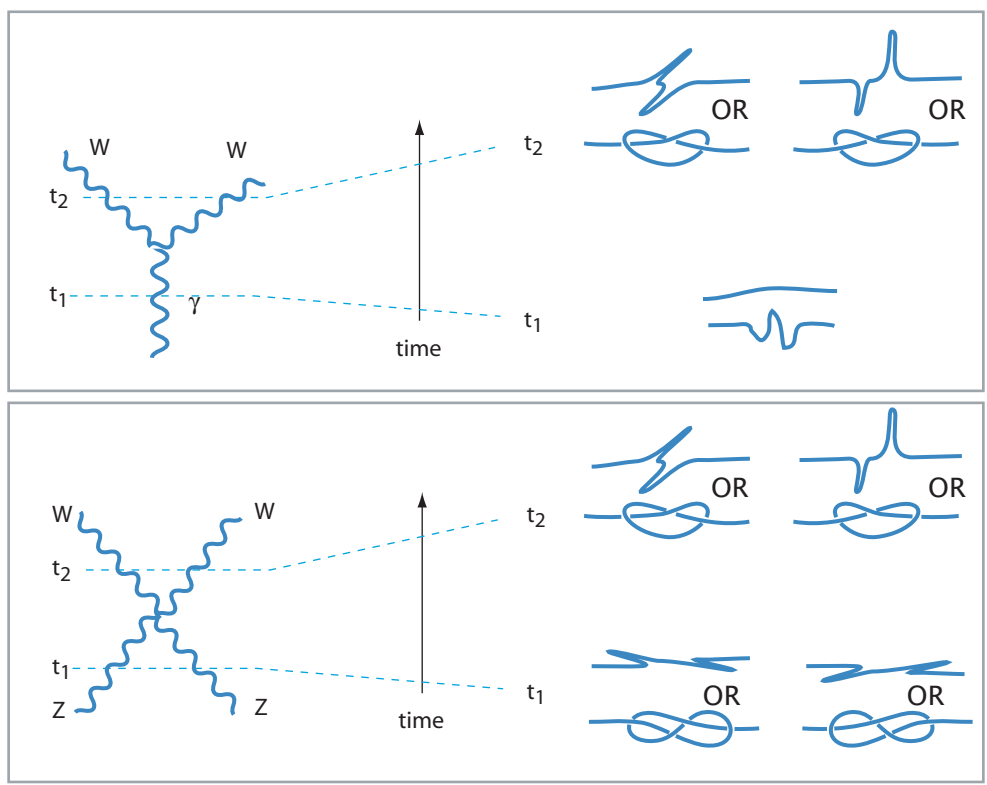
\csepsf{i-ee-ternary}{scale=1}{Some ternary and quartic boson couplings in the
weak interaction.}

Some ternary and quartic couplings among the vector bosons are shown in
\figureref{i-ee-ternary}.  The strand model reproduces these
processes.

In all these weak strand reactions, total electrical charge, total weak
isospin, baryon number and total spin are conserved, as is indeed observed.
In this way, the tangle model for fermions reproduces
the weak interaction, provided that the $\rm Z$ and $\rm W$ bosons are each
made of one knotted strand.

%-----------------------------------------------------------------------------
% Feb 2009 
\subsection{Renormalization of the weak interaction}

The strand model with its limited number of strands that appear in elementary
particle reactions, together with the equivalence of pokes and the weak
interaction, implies that only triple and quadruple vertices are possible;
higher order vertices are impossible in the strand model.  This is the central
requirement for the renormalization of the theory.  The strand model also
reproduces the experimentally verified terms of the electroweak Lagrangian.
(Quark and fermion mixing, as well as the Higgs boson, are discussed elsewhere
\cite{cs5}.)  The strand model thus automatically ensures that the electroweak
interaction is renormalizable.
% 
% The strand model thus provides a new underlying picture for the Feynman
% diagrams of the electroweak interaction, but does not change the physical
% results at any experimentally accessible energy scale.  
In particular, the
running of the weak coupling constant and of the masses of particles are
reproduced.

%-----------------------------------------------------------------------------
% Feb 2009 
\subsection{Deviations from the standard model}

The strand model suggests that there are no deviations from the quantum field
theory of the electroweak interaction for any experimentally accessible
energy.  Only when gravity is of importance, the strand model predicts a
maximum electroweak field, given by the Planck value.  Neutron stars, quarks
stars, gamma ray bursters, quasars and all other astrophysical phenomena are
predicted to have field values below the limit value. So far, no observed 
value violates the predicted limit.

The strand model might deviate form the electroweak interaction only at Planck
scales.  No other gauge group comes into play even at highest energies.  In
particular, the strand model predicts the absence of larger gauge groups, such
as those as conjectured by grand unification.

% \end{document} bombs
% \endinput

%-----------------------------------------------------------------------------
% Feb 2009, March 2009
\section{The strong interaction and its SU(3) gauge group}

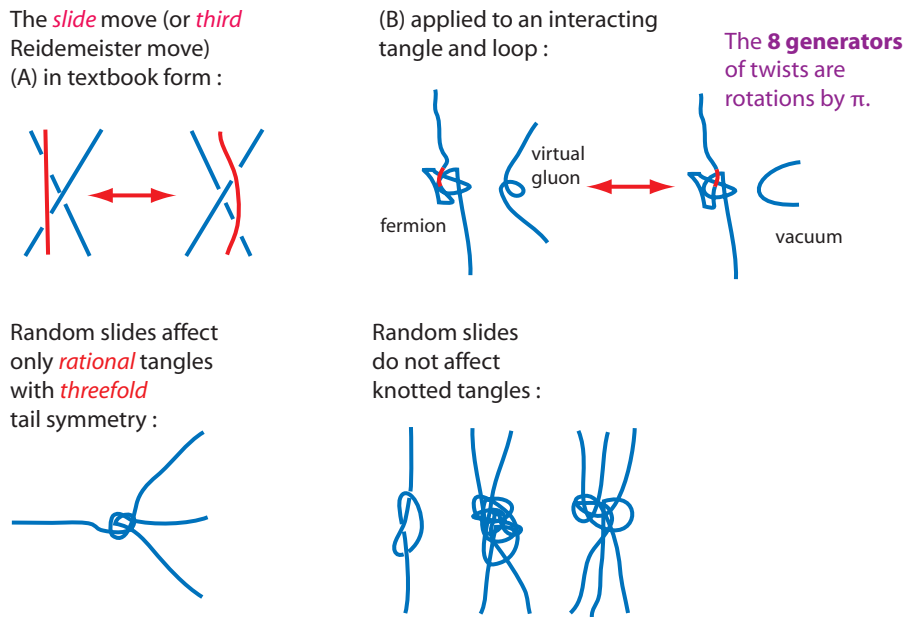
\csepsf{su3}{scale=0.75}{\emph{Slides} as tangle deformations, a
tangle that is affected by slides, and some tangles that are not.}

The third way to deform a tangle core while changing crossing numbers is to
\emph{slide} a piece of strand over a crossing of two other pieces, as shown
in \figureref{su3}.  In knot theory, this type of deformation is called the
\emph{third Reidemeister move}.  We will see shortly that the slide
deformation of a tangle is related to what is usually called the emission or
absorption of a gluon.

Certain tangle cores will be affected by large numbers of random slides
whereas others will not be.  For example, if we take a \emph{single}, knotted
strand, we see that it will be unaffected.  In other words, single knotted
strands do not interact strongly; indeed, the W and the Z are observed to be
`white' in the terminology of the colour force.  Also the unknotted photon is
predicted to be `white' and thus to transform under a singlet representation.
The same happens for all fermions that are prime tangles of \emph{three}
strands: the strand model predicts \cite{cs5} that leptons do not interact
strongly, as indeed is observed.

In fact, only few tangles are affected by large numbers of random slides.  The
tangle given in the lower left of \figureref{su3}, with 4 tails, is an
example.  We note that in this case, the 4-tailed tangle of \figureref{su3}
transforms as a triplet representation.  (Thus we can call each of the three
possible options `coloured'.)  In fact, elementary tangle cores with mass that
transform following other -- e.g., \emph{faithful} -- representations of SU(3)
are impossible.
% The strand model
% thus recovers the result that no elementary fermion transforms under a
% faithful SU(3) representation.  
In other words, the strand model states that the photon and all massive
elementary particles are either singlet (`neutral') or triplet (`coloured')
representations of SU(3), as is observed.

% \end{document} % bombs
% \endinput

%-----------------------------------------------------------------------------
% Feb 2009, March 2009
\subsection{Gluons, SU(3) and QCD}

% \csepsf{i-su3gauge}{scale=1}{Slides generate an SU(3) group (see text),
% as can be seen when the three planes containing the generator axes are modelled as
% three belts with buckles that can tilt around connecting joints.}

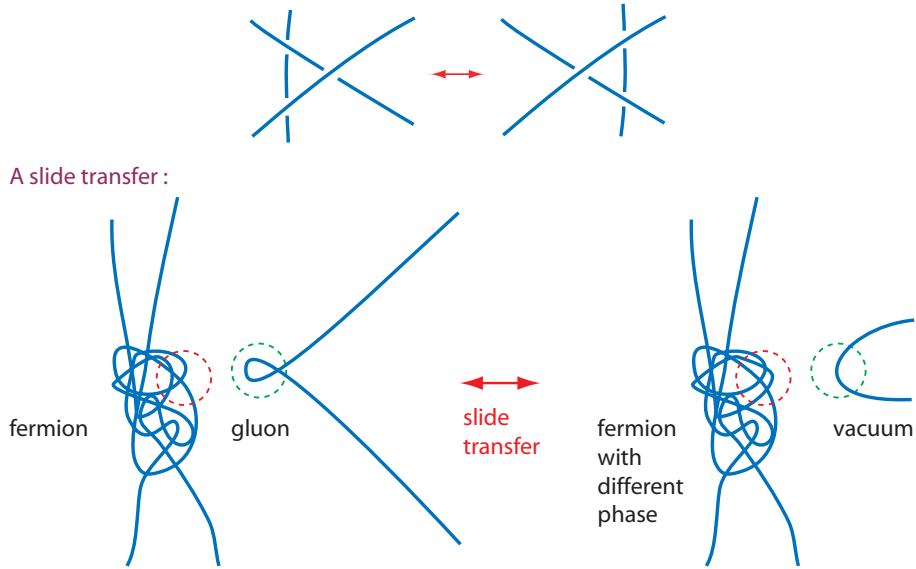
\csepsf{i-strands-strong}{scale=1}{A single gluon strand changes the
rotation of a tangle: \emph{slide transfer} is the basis of the strong
interaction in the strand model.  No strand is cut or reglued; the transfer
occurs only through the excluded volume due to the impenetrability of
strands.}

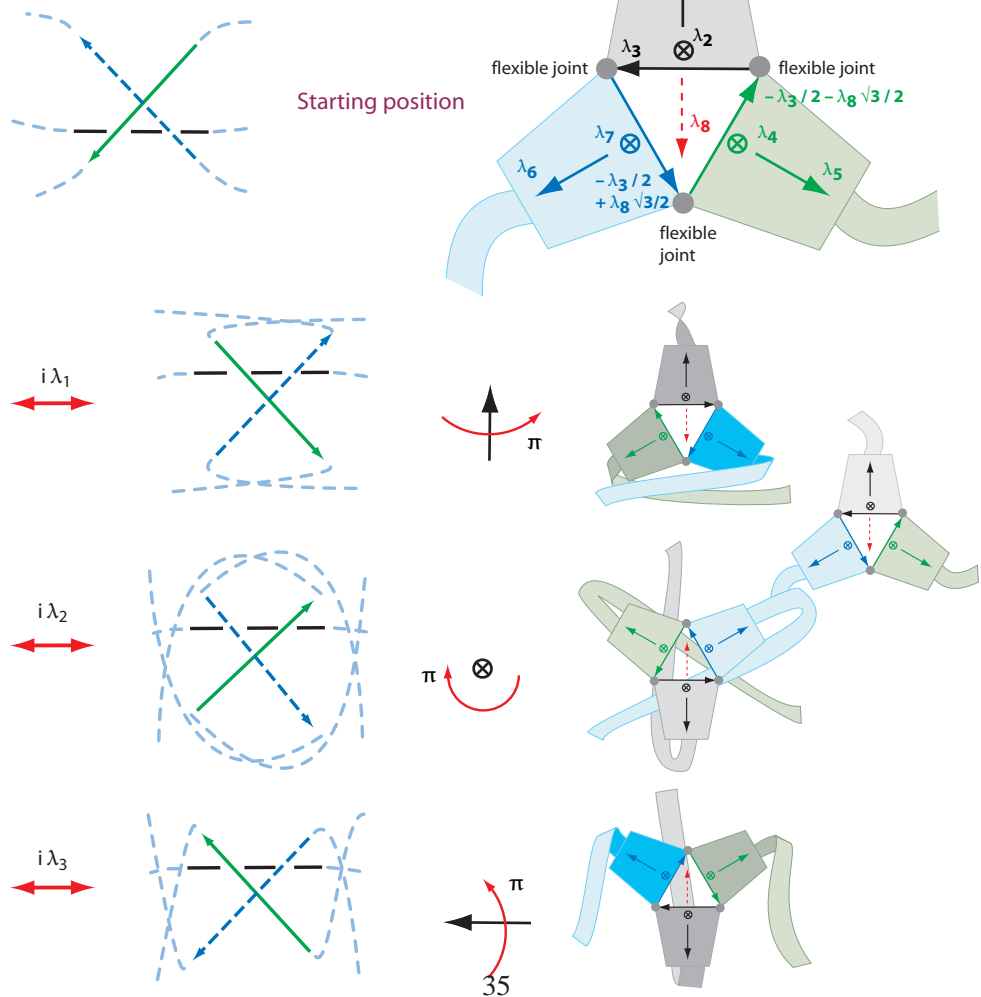
\csepsf{i-su3gauge}{scale=1}{The strand deformations for the eight gluons
$\lambda_{1}$ to $\lambda_{8}$ associated to the slide move, and the SU(3)
structure of the strong interaction as the result of three joined belts.}

The third Reidemeister move involves three pieces of strands; the move deforms
a tangle core by sliding a crossing over or under a third strand, as shown in
\figureref{i-strands-strong}.  In three dimensions, this operation can be
realized in several ways, called $\lambda_{0}$ to $\lambda_{3}$ and shown in
\figureref{i-su3gauge}.  The `naive' slide $\lambda_{0}$ involves no crossing
switch and is thus unobservable in the strand model.  The slides
$\lambda_{1}$, $\lambda_{2}$ and $\lambda_{3}$ involve combined rotations by
$\pi$ of two strands, thus involve crossing switches, and therefore are
physical.  We note that `slide' is not a good term for these operations; in
fact, they are combinations of a \emph{rotation} by $\pi$ and a
\emph{flattening} into the observation plane.  Nevertheless, we will continue
to call them `slides' for brevity.

We can find a visualization of SU(3) if we imagine three belts whose buckles
are attached at joints, as illustrated in \figureref{i-su3gauge}.  Three
slides are attached to each buckle, thus leading to 9 slides in total.  (The
slides corresponding to $\lambda_{1}$ are usually called $\lambda_{5}$ and
$\lambda_{6}$, those to $\lambda_{2}$ are called $\lambda_{4}$ and
$\lambda_{7}$.)  Three of the slides constructed in this way are linearly
dependent, namely those formed by $\lambda_{3}$ and its two `cousins'; among
them, only two are needed (called $\lambda_{3}$ and $\lambda_{8}$ in the
Gell-Mann set), giving a total of 8 slides.  The slides $\lambda_{1}$,
$\lambda_{2}$ and $\lambda_{3}$ form an SU(2) subgroup, and the same happens
on the other buckles.  This models the three linearly independent SU(2)
subgroups contained in SU(3).

The definition of the 8 slides allows them to be concatenated.  To explore
concatenations, a few details are important.  First, the slides of
\figureref{i-su3gauge} correspond to $i$ times the Gell-Mann generators.
Second, the slide $\lambda_{8}$ that makes $\lambda_{9}$ unnecessary is
orthogonal to $\lambda_{3}$.  Third, the triplet $\lambda_{1}$, $\lambda_{2}$,
$\lambda_{3}$ forms a SU(2) subgroup, as does the triplet $\lambda_{5}$,
$\lambda_{4}$, $-\lambda_{3}/2 - \lambda_{8} \sqrt{3}/2$ and the triplet
$\lambda_{6}$, $\lambda_{7}$, $-\lambda_{3}/2 + \lambda_{8} \sqrt{3}/2$.
Fourth, all slides are combinations of rotations and flattenings.  For this
reason, their square is not $-1$, but involves $\lambda_{8}$ and/or
$\lambda_{3}$.  Fifth, multiplying slides is concatenation, whereas adding
slides is an operation defined in \cite{cs3}.  Loosely speaking, addition
connects partial tangles without additional crossings.
% Sixth, when slides from the same triplet

% Two general slides do not commute and do not anticommute.  Closer
% inspection shows that their commutation relations (note that there is
% a factor $i$ the definition) turn out to be:
% 
% \medskip
% 
\def\s{\sqrt{3}}\let\l\lambda 

\normalsize
The multiplication table is:
\footnotesize
$$
\hspace*{-12mm}\begin{array}{|c||c|c|c||c|c||c|c|c|} \hline
  \cdot      & \l_1    &    \l_2 &    \l_3 &    \l_4 &    \l_5 &    \l_6 &    \l_7 &    \l_8     \\ 
\hline \hline
\l_1    &   2/3   &  i\l_3 & -i\l_2 &   i\l_7/2 &  -i\l_6/2 &   i\l_5/2 &  
-i\l_4/2 &           \\ 
      &+ \l_8/\s&          &          &   +\l_6/2 &   +\l_7/2  &  +\l_4/2   
      &   +\l_5/2  &  +\l_1/\s \\ 
      \hline
\l_2    & -i\l_3 &     2/3 &  i\l_1 &   i\l_6/2 &   i\l_7/2 &  -i\l_4/2 &  -i\l_5/2 
&           \\ 
      &         &+ \l_8/\s&           &   -\l_7/2 &   +\l_6/2  &   +\l_5/2  &   
      -\l_4/2  &  + \l_2/\s \\ 
\hline
\l_3    &  i\l_2 & -i\l_1 &       2/3 &   i\l_5/2 &  -i\l_4/2 &  -i\l_7/2 &   
i\l_6/2 &          \\ 
      &          &          &+ \l_8/\s&   +\l_4/2 &   +\l_5/2  &   -\l_6/2  &  
      -\l_7/2  &  +\l_3/\s \\ 
\hline
\hline
\l_4    &  -i\l_7/2 &  -i\l_6/2 &  -i\l_5/2 &  2/3+\l_3/2 &   i\l_3/2 &   i\l_2/2 &   i\l_1/2 & -i\s\l_5/2 \\
      & +\l_6/2   &  -\l_7/2  &  + \l_4/2 & -\l_8/2\s & +i\s\l_8/2  &   +\l_1/2 &   
      -\l_2/2 & - \l_4/2\s \\ 
\hline
\l_5    &  i\l_6/2  &  -i\l_7/2 &   i\l_4/2 &  -i\l_3/2 & 2/3+\l_3/2  &  -i\l_1/2 &   i\l_2/2 & i\s\l_4/2 \\
      & +\l_7/2    &  +\l_6/2   &    +\l_5/2 & -i\s\l_8/2 & -\l_8/2\s &   +\l_2/2 &  
      +\l_1/2   &   - \l_5/2\s  \\ 
\hline
\hline
\l_6    & -i\l_5/2  &   i\l_4/2 &   i\l_7/2 &  -i\l_2/2 &   i\l_1/2 &  2/3-\l_3/2 &  -i\l_3/2 &-i\s\l_7/2 \\ 
      & +\l_4/2   &  +\l_5/2  &   -\l_6/2 &  +\l_1/2     &   +\l_2/2  & 
      -\l_8/2\s & +i\s\l_8/2 &  - \l_6/2\s\\ 
\hline
\l_7    &  i\l_4/2  &   i\l_5/2 &  -i\l_6/2 &  -i\l_1/2 &  -i\l_2/2 &   
i\l_3/2 & 2/3-2\l_3 & i\s\l_6/2 \\ 
      &  +\l_5/2   &  -\l_4/2  &   -\l_7/2 &  -\l_2/2     &   +\l_1/2  &  -i\s\l_8/2 &  
      -\l_8/2\s  & - \l_7/2\s\\ 
\hline
\l_8    &         &         &        
&i\s\l_5/2&-i\s\l_4/2&i\s\l_7/2&-i\s\l_6/2& 2/3 \\ 
      &+ \l_1/\s&+ \l_2/\s&+ \l_3/\s&- \l_4/2\s&-  \l_5/2\s&-  \l_6/2\s&-  
      \l_7/2\s&  -\l_8/\s \\ 
\hline
\end{array}
$$

\medskip

% \noindent\normalsize These tables are those of the Gell-Mann matrices, which
\noindent\normalsize This table is precisely that of the Gell-Mann matrices,
which form a standard set of generators of the group SU(3).  We thus conclude
that the eight linearly independent generalized slides that can be applied to
a tangle represent virtual gluons that act on a particle.

The slide analogy for \emph{virtual} gluons implies that \emph{real} gluons
are described by single unknotted strands that impart `slides' to fermions.  A
simple image is to describe real gluons as \emph{loops} that `pull' one strand
during the slide, as shown in \figureref{i-ee-gluons}.  This single strand
model also reproduces the vanishing mass of gluons and their spin 1 value.

The 8 gluons transform according to the adjoint (and faithful) representation
of SU(3).  The model for gluons implies that two interacting gluons can yield
either one or two gluons, but not more, as shown in
\figureref{i-ee-qcdtriple}.  Since in the strand model, gluons do not change
topology, but only shapes, gluons are predicted to be massless, despite
interacting among themselves.  In total, after averaging over space and time,
we thus get the usual free gluon Lagrangian \begin{equation}{L}_{\mathrm{gluons}} = -
\frac{1}{4} G^g {}_{\mu\nu} G_g {}^{\mu\nu}\end{equation}
 from the strand model.  

A structure made of two or three gluons would not be knotted or linked.  It is
unclear whether such a structure is stable in the strand model, though
appearance seems to speak against the idea.  Therefore it seems that glueballs
might not exist.  Despite this apparent lack of a mass gap, the lack of
classical gluonic waves is explained by the triple and quartic gluon vertices.

In the strand model, `colour' is the name give to the property distinguishing
the three states forming triplet representations.  Simple visual inspection
shows that slides, and thus gluons, \emph{conserve} colour.

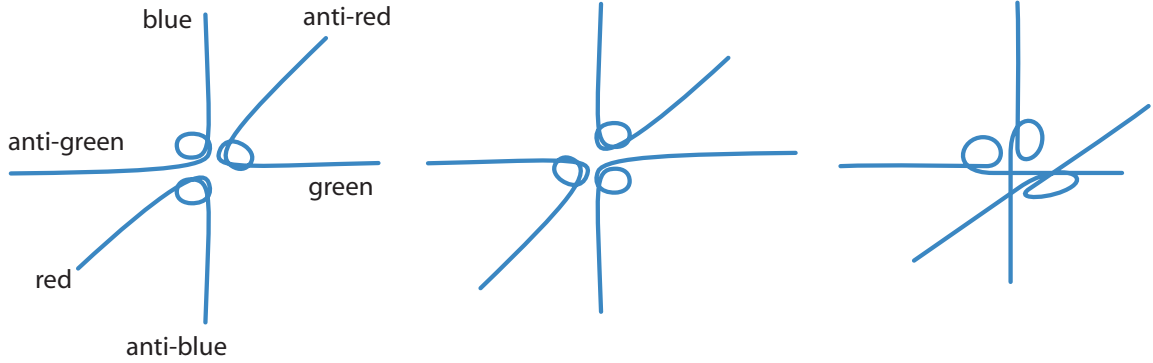
\csepsf{i-ee-gluons}{scale=1}{The strand model for the nine {gluons}, the last
three not being linearly independent.}

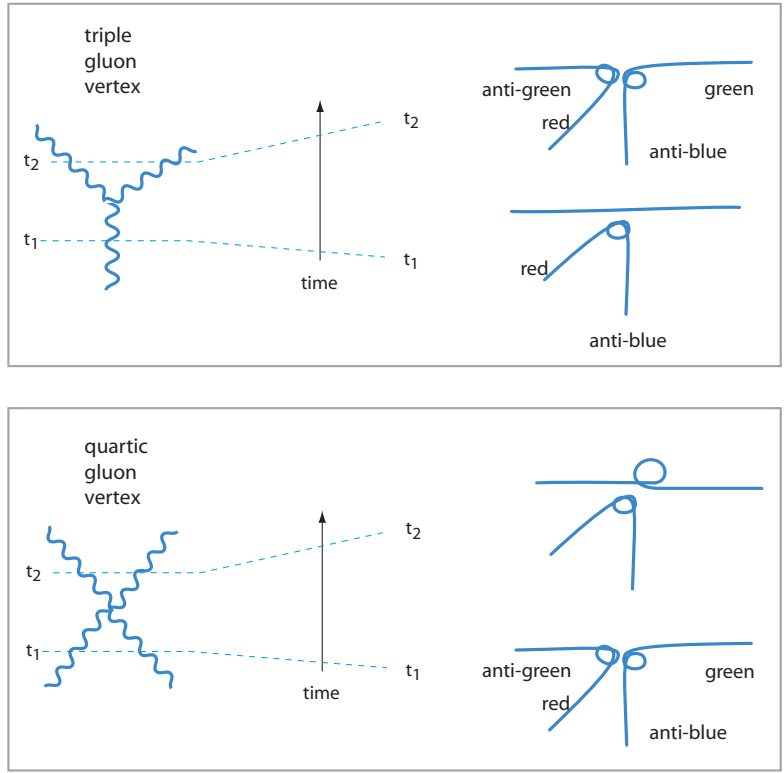
\csepsf{i-ee-qcdtriple}{scale=0.8}{The self-interaction of gluons in the
strand model.}

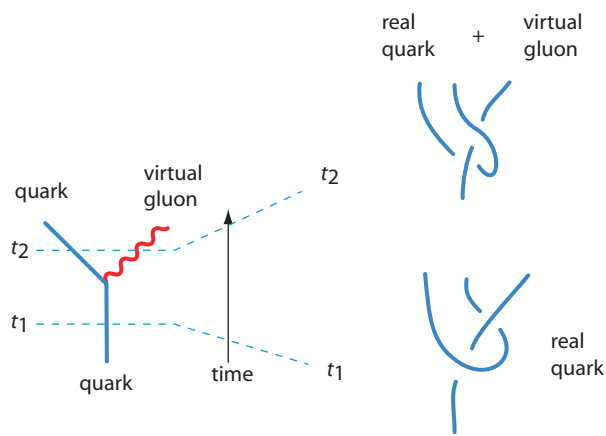
\csepsf{i-ee-qcd}{scale=1}{The Feynman diagram of the strong interaction for a
quark.}

Slides, i.e., gluon emission or absorption, never change the topology of
tangles.  Thus the strand model predicts that the strong interactions conserve
electric charge, baryon number, weak isospin, flavour, spin and all parities.
This is indeed observed. In particular, there is a natural lack of $CP$
violation by slides.  This is exactly what is observed about the strong
interaction.  The strand model thus also reproduces the lack of $CP$ violation
by the strong interaction.

In short, in the strand model, the emission or
absorption of a virtual gluon by a quark is expected to be a rearrangement of
the strand or tails of a rational tangle, as shown in \figureref{i-ee-qcd}.
The specific tangles for all quarks and hadrons, with the resulting
properties, are discussed elsewhere \cite{cs5}.

Starting from the idea that tangle core deformations lead to phase
redefinitions, we have thus found that generalized slides imply an SU(3)
invariance.  In other words we find that the complete strong interaction
Lagrangian density for matter and radiation fields is SU(3) gauge invariant.
After averaging crossings over space and time we thus get the well-known
Lagrangian \begin{equation}{L}_{\mathrm{QCD}} = \sum_q \bar \psi_q (i \hbar
c\displaystyle{\not}D - m_q c^2) \psi_q - \frac{1}{4} G^g {}_{\mu\nu} G_g
{}^{\mu\nu}\end{equation} where $\displaystyle{\not}D$ is now the SU(3)
gauge-covariant derivative.  In short: the strand model reproduces QCD. Only
the number of quarks, the values of their masses, and the coupling strength
remains open.

%-----------------------------------------------------------------------------
% Apr 2008
\subsection{Quark confinement}

In the strand model, rational tangles represent quarks.  Rational tangles are
topologically unstable; thus they are not localizable and the do not behave
like or represent free particles.  In this way, the strand model explains the
lack of free quarks.  The strand model also explains the lack of quark decay,
as explained elsewhere \cite{cs5}.

In the strand model, a bond between two quarks is a structure that connects
the tails of the quarks.  The simplest case is that of mesons, where a quark
bonds to an antiquark.  In this case, three strands form the bond between the
quark and the antiquark, thus realizing something similar to the original
`hadronic string model'.  Since the resulting effective potential in mesons is
confined to a tube, the strand model reproduces Regge trajectories, their
common slope, mass sequences, quantum numbers, flavour oscillations, and all
other properties of mesons.  Also baryons are modelled successfully.  More
details and comparisons with experimental data are given elsewhere \cite{cs5}.

%-----------------------------------------------------------------------------
% Feb 2009 
\subsection{Renormalization of the strong interaction}

The strand model, together with the slide move, implies that only a limited
number of Feynman diagrams appear in strong nuclear reactions.  In particular,
the slide move implies that only one QCD Feynman diagram exists for quarks,
and that only triple and quadruple vertices exist for gluons.  This limited
range of options is essential for the renormalization of QCD and allowed to
deduce the QCD Lagrangian.  The strand model thus ensures that the strong
interaction is renormalizable.

The strand model provides a new underlying picture for the Feynman diagrams of
the strong interaction, but does not change the physical results at any energy
scale accessible in the laboratory.  In particular, the running of the strong
coupling constant is reproduced.  (In the strand model, a flux-tube--like bond
between the quarks appears automatically \cite{cs5}.  At high kinetic
energies, the bond has little effect, so that quarks behave more like free
particles.  The strand model thus reproduces asymptotic freedom.)

%-----------------------------------------------------------------------------
% Feb 2009
\subsection{Deviations of the strand model from QCD}

The strand model thus seems to reproduce QCD in its essential aspects: gauge
symmetry of the Lagrangian, asymptotic freedom, and quark confinement.
Deviations from QCD are thus only expected near the Planck scale, when the
spatial and temporal averaging of crossing switches is not possible.  Again,
the Planck value for the gluon field intensity is predicted to be the highest
possible field value.  Neutron stars, quark stars and all other astrophysical
objects are predicted to have field values below this limit.  This is
observed.  In short, deviations of the strand model from QCD are not expected
for any laboratory energy.

On the other hand, the strand model implies that no other gauge group other
than SU(3) comes into play, even at high energy; in particular, the strand
model again predicts the absence of grand unification.

%-----------------------------------------------------------------------------
% Feb 2009
\section{Lack of other interactions}

Already in 1926, Kurt Reidemeister proved an important theorem about possible
deformations of knots or tangles that involve crossing switches.  When tangles
are described with two-dimensional diagrams, all possible deformations can be
reduced to \emph{exactly three moves}, nowadays called after him
\cite{reidem}.  In the strand model, the two-dimensional tangle diagram is
what an observer \emph{observes} about a physical system.  Reidemeister's
theorem, together with the equivalence of interactions as crossing-switching
deformations, thus proves that there are \emph{only three interactions} in
nature.

On the other hand, it is well known that in fact, there is only \emph{one}
Reidemeister move \cite{lou}.  This is especially clear if one looks at the
three-dimensional shape of knots or tangles, instead of at their
two-dimensional diagrams.  In the terms of the strand model, this means that
all interactions are in fact aspects of only \emph{one} basic process.  Indeed, the
strand model asserts that all interactions are consequences of strand
deformations (themselves due to fluctuations).  In this way, the three
interactions are thus unified by the strand model.

In other words, the strand model asserts that there is no single gauge group
for all gauge interactions.  The predicted absence of grand unification
implies the absence of large proton decay rates, the absence of new gauge
bosons, and the absence of large electric dipole moments in elementary
particles.

Furthermore, the strand model uses only three spatial dimensions and
has no evident supersymmetry.  In this sense, the strand model differs
from many unification proposals made in the past decades.  This allows
us to draw some interesting conclusions.

%-----------------------------------------------------------------------------
% Jan 2007
\section{Coupling constants}

In the strand model, unification of gauge interactions does not lead
to a common, unique gauge group at high energies.  In other words, unification
in the strand model does \emph{not necessarily} mean that the three
coupling constants must converge to a unique value.  In fact, the
strand model suggests that the three coupling constants will be
related, but not by an equality at any particular energy.

First of all, the strand model suggests that the coupling constants are
\emph{calculable}.  Indeed, coupling constants give the probability with which
virtual or real bosons are emitted or absorbed by charged particles.  For a
straightforward calculation of the coupling constant based on the strand
model, we can simply determine the probability of the relevant virtual gauge
boson deformations when a charged tangle fluctuates.  Computer simulations are
the calculation method of choice.

Relating coupling constant to probabilities of core deformations  means
that all coupling constants are smaller than 1, for all energies.  This is
observed.
The strand model also predicts that the coupling constants are constant in
time, as the mechanism at the basis of gauge interactions does not depend on
the size of the universe.

Furthermore, the strand model predicts that all three coupling constants are
independent of the specific tangle topology, as long as the relevant charge
value is the same.  This strand model prediction can also be tested with
computer simulations.

Additional checks are possible.  Using a sufficiently large
statistical basis, we can  simulate the collision between a fermion and
a boson, or the fluctuations around a charged fermion, or the collision of two
charged particles.  All these methods must yield the same result for each
coupling constant; this provides a consistency check of the strand model.

Calculating coupling constants with statistical simulations will require large
computer processing time.  The existing software packages developed for
polymer simulation, for cosmic string evolution and for vortex motion in
superfluid materials might be the best candidates to perform such
calculations.

In summary, the strand model suggest a way to calculate coupling constants.
Checking the numerical correspondence will provide a definite test for the
model.

%-----------------------------------------------------------------------------
\section{Is the strand model the simplest possible?}

In order to reproduce three-dimensional space, Planck units, spin, and
black-hole entropy, extended fundamental entities are required. % \cite{cs2,cs3}.
Compared to the strand model, other models based on other extended objects --
such as bands, strings, membranes, ribbons, posets, branched lines, networks,
crystals or virtual knots -- increase the complexity in two ways: these models
add \emph{features} to the fundamental entities and they complicate the
\emph{mapping} from the model to observation.

In many models, the fundamental entities have (additional) \emph{features}.
Examples are ends \cite{wen}, width or twists \cite{a,sb}, field values \cite{rjf2007},
coordinates, quantum numbers \cite{smo},
% \cite{s},
tension, or non-trivial topological information \cite{loulo,posets,davidfin}.
However, any added feature increases the complexity of the model and also is
an assumption that is difficult to justify.  In contrast, the strand model
uses featureless entities and has less justification issues.
% (Still
% other models, such as non-commutative space-time or supergravity, do not seem
% to use common constituents for matter and vacuum.)

Secondly, the \emph{mapping} between the more complex models and experiment is
often {intricate} and sometimes not unique.  In contrast, the strand model
argues that the experimentally accessible Dirac equation, and thus quantum
field theory, and the experimentally accessible field equations of general
relativity arise \emph{directly} from Planck scales, through an averaging
procedure.  In this way, the strand model proposes to unify the two halves of
physics with only one additional postulate: strand crossing switches define
Planck units.  In particular, the strand model proposes that not only vacuum
and matter, but also gauge interactions are natural consequences of the strand
structure of particles and space at Planck scales.  The comparable ideas in
other models are much more elaborate.

Therefore the strand model might be the unified model with the smallest number
of additional concepts, thus satisfying Occam's razor.

%-----------------------------------------------------------------------------
\section{Is the strand model a unified description?}

Any unified description of nature must first of all provide a precise
description of \emph{all} observations.  This can only be tested by
experiment.  So far, the model has not been falsified.  Many new experiments
designed for further tests, including the LHC in Geneva, will provide
additional tests.  But a unified description must also have an additional
property: it must be \emph{unmodifiable}.  A unified description must leave no
alternative.

If a unified description can be modified, it loses its explanatory power.
(David Deutsch states that any good explanation must be `hard to vary'
\cite{ted}.)  In particular, the requirement means that a unified description
must be impossible to generalize, and that it must be impossible to reduce the
unified description to special cases.  Exploring the strand model shows that
it fulfils these conditions: in particular, the strand model does not work for
other spatial dimensions, for other types of fundamental entities, or for
other definitions of the Planck units \cite{cs5}.

Therefore, the strand model is a candidate for a unified description -- as
long as it is not falsified by experiment.

%-----------------------------------------------------------------------------
\section{Outlook}

\noindent In short, a model of particles based on fluctuating featureless
tangled strands in three spatial dimensions appears to yield the gauge
interactions of the standard model in a natural way.  The electromagnetic,
weak and strong interaction behaviour of fermions and bosons appears to follow
naturally from tangle core deformations due to the first, second and third
Reidemeister move.  The U(1), SU(2) and SU(3) gauge groups, parity violation
of the weak interaction, SU(2) symmetry breaking, and asymptotic freedom of
the strong interactions appear to be reproduced in a natural way.  

The strand model yields most observed terms of the standard model Lagrangian,
and no contradictions with experiments appear.  The model also proposes a way
to calculate coupling constants.  Such a calculation will allow the definitive
test of the model.

A separate investigation \cite{cs5} clarifies the issues left open by the the
present results, in particular, it clarifies the origin of the three
generations, the precise tangles for each elementary particle, their masses,
their mixing matrices, as well as the properties of the Higgs boson.

The strand model also suggests that the standard model of particle physics is
valid up to high energies.  Maximum field values for all gauge fields are
predicted.  The strand model predicts the absence of $CP$ violation for the
strong interaction and thus the absence of axions.  The model also predicts
the absence of grand unification, of the corresponding vector bosons, of
supersymmetry, of supersymmetric partner particles, of higher spatial
dimensions, and of all the experimental effects associated with them, such as
new decays, new reactions, or large electric dipole moments.  Glueballs
probably do not to exist.

In summary, the strand model, a fully algebraic model of fundamental physics,
reproduces the standard model of particle physics, quantum theory, and general
relativity, while not allowing any alternative or extension.
%

%-----------------------------------------------------------------------------
\section{Acknowledgments}

I thank {Louis Kauffman}, {Claus Ernst}, {Andrzej Stasiak} and Warren Siegel
for extensive discussions.  I also thank {Greg Egan} for extending his online
applet to show both ways to realize the belt trick.  And I thank Jack Avrin,
Bill Baylis, Thomas Racey, Garnet Ord, Peter Battey Pratt, Michael Berry,
Sundance Bilson-Thompson, Luca Bombelli, Steve Carlip, David Finkelstein, Ted
Jacobson, Oren Raz, Lee Smolin, Frank Wilczek, Jerrold Marsden, Alden Mead,
Alfred Shapere, and Thanu Padmanabhan for feedback and suggestions.

%-----------------------------------------------------------------------------


\begin{thebibliography}{99}


\bibitem[Avrin 2005]{a} J.S. Avrin, \emph{A visualizable representation of the
elementary particles}, {Journal of Knot Theory and Its Ramifications} {\bf
14}, pp.~{131--176} (2005).
    
\bibitem[Baylis \& al. 2007]{baylistt}
W.E. Baylis, R. Cabrera, D. Keselica \emph{Quantum\slash classical interface:
fermion spin}, preprint at \url{arxiv.org/abs/0710.3144}.

\bibitem[Battey-Pratt \& Racey 1980]{bpr} E. Battey-Pratt and T.
Racey, \emph{Geometric model for fundamental particles}, International
Journal of Theoretical Physics {\bf 19}, pp.  437--475 (1980).

% \bibitem[Baylis 2004]{baylis}
% W.E.~Baylis, \emph{Surprising symmetries in relativistic charge
% dynamics}, preprint at \url{arxiv.org/abs/physics/0410197}.

\bibitem[Berry 1984]{berry}
M.V. Berry, \emph{Quantal phase factors accompanying adiabatic changes},
Proceedings of the Royal Society A, {\bf 392}, pp.  45--57 (1984).

\bibitem[Bilson-Thompson \& al.  2005-2008]{sb} S. Bilson-Thompson, \emph{A
topological model of composite preons}, \url{arxiv.org/hep-ph/0503213}; S.
Bilson-Thompson, F. Markopoulou and L. Smolin, \emph{Quantum gravity and the
standard model}, \url{arxiv.org/hep-th/0603022}; S. Bilson-Thompson, J.
Hackett, L. Kauffman and L. Smolin, \emph{Particle identifications from
symmetries of braided ribbon network invariants},
\url{arxiv.org/abs/0804.0037}; S. Bilson-Thompson, J. Hackett, and L.H.
Kauffman, \emph{Particle topology, braids, and braided belts},
\url{arxiv.org/abs/0903.1376}.

% \bibitem[Bilson-Thompson 2005]{sb1} S. Bilson-Thompson, \emph{A
% topological model of composite preons},
% \url{arxiv.org/hep-ph/0503213}.
% 
% \bibitem[Bilson-Thompson et al. 2006]{sb2} S. Bilson-Thompson, F. 
% Markopoulou and L.
% Smolin, \emph{Quantum gravity and the standard model},
% \url{arxiv.org/hep-th/0603022}.
% 
% \bibitem[Bilson-Thompson et al.~2008]{sb3} S. Bilson-Thompson, J.
% Hackett, L. Kauffman and L. Smolin, \emph{Particle identifications from
% symmetries of braided ribbon network invariants},
% \url{arxiv.org/abs/0804.0037}. 

\bibitem[Bohm \& al.  1955]{bstiomno} D. Bohm, R. Schiller, J. Tiomno, \emph{A
causal interpretation of the Pauli equation (A)}, Supplementi al Nuovo Cimento
{\bf 1}, pp.  48--66 (1955).

\bibitem[Bombelli \& al.  1987]{posets} L. Bombelli, J. Lee, D. Meyer
and R. Sorkin, \emph{Space-time as a causal set}, Physical Review
Letters \textbf{59}, pp.  521--524 (1987). See also the review by J.
Henson, \emph{The causal set approach to quantum gravity},
\url{arxiv.org/abs/gr-qc/0601121}.

\bibitem[Carlip 2010]{cup} S. Carlip, \emph{The small scale structure of
spacetime}, preprint at \url{arxiv.org/abs/1009.1136}.

\bibitem[Deutsch 2009]{ted} D. Deutsch, \emph{A new way to explain
explanation}, talk at \url{www.ted.org}.

\bibitem[Egan 2009]{egan} G. Egan, animation applet on
\url{www.gregegan.net/APPLETS/21/21.html}.

\bibitem[Feynman 1988]{feynqed} R.P. Feynman, \emph{QED -- The Strange
Theory of Light and Matter}, Princeton University Press (1988).

\bibitem[Finkelstein 2008]{davidfin} D. Finkelstein, \emph{Homotopy
approach to quantum gravity}, International Journal of Theoretical
Physics \textbf{47}, pp.  534--552, (2008).

\bibitem[Finkelstein 2007]{rjf2007} R.J. Finkelstein, \emph{A field theory of
knotted solitons}, \url{arxiv.org/abs/hep-th/0701124}. See also
%\bibitem[Finkelstein 2006]{rjf2006} 
R.J. Finkelstein, \emph{Trefoil solitons,
elementary fermions, and $\rm SU_{\rm q}(2)$}, 
\url{arxiv.org/abs/hep-th/0602098};
%
%\bibitem[Finkelstein 2005]{rjf2005} 
R.J. Finkelstein and A.C. Cadavid, \emph{Masses and
interactions of q-fermionic knots}, 
\url{arxiv.org/abs/hep-th/0507022};
%\bibitem[Finkelstein 2004]{rjf2004} 
R.J. Finkelstein, \emph{A knot model
suggested by the standard electroweak theory},
\url{arxiv.org/abs/hep-th/0408218}.

% \bibitem[Gottfried \& Weisskopf 1984]{weisskopf} K. Gottfried and V. Weisskopf, 
% \emph{Concepts of Particle Physics}, volume 1 (Clarendon Press, 1984).

\bibitem[Jacobson 1995]{jacobgg} T. Jacobson, \emph{Thermodynamics of
spacetime: the Einstein equation of state}, {Physical Review Letters} {\bf
75}, pp.  1260--1263 (1995), preprint at
\url{www.arxiv.org/abs/gr-qc/9504004}.

\bibitem[Kauffman 1991]{lou} L.H. Kauffman, \emph{Knots and Physics},
World Scientific (1991).

\bibitem[Kauffman \& Lomonaco 2004]{loulo} L.H. Kauffman and S.J.
Lomonaco, \emph{Quantum knots}, \url{arxiv.org/abs/quant-ph/0403228}.
See also S.J. Lomonaco and L.H. Kauffman, \emph{Quantum knots and
mosaics}, \url{arxiv.org/abs/quant-ph/0805.0339}.

% \bibitem[Ng 2002]{s} S.K. Ng, \emph{On a knot model of the $\pi^+$
% meson}, \url{arxiv.org/abs/hep-th/0210024}, and \emph{On a
% classification of mesons}, \url{arxiv.org/abs/hep-ph/0212334}.

\bibitem[Padmanabhan 2009]{pad} T. Padmanabhan, \emph{Equipartition of energy
in the horizon degrees of freedom and the emergence of gravity},
\url{arxiv.org/abs/0912.3165}.

\bibitem[Putterman \& Raz 2008]{squarecat} E. Putterman and O. Raz,
\emph{The square cat}, American Journal of Physics {\bf 76},
pp.~1040--1045 (2008).

\bibitem[Reidemeister 1926]{reidem} K. Reidemeister, \emph{Elementare
Begründung der Knotentheorie}, Abhandlungen aus dem Mathematischen
Seminar der Universität Hamburg \textbf{5}, pp.  24--32 (1926).
    
\bibitem[Schiller 2005]{cs2a} C. Schiller, \emph{General relativity and
cosmology derived from principle of maximum power or force}, International
Journal of Theoretical Physics {\bf 44}, pp.  1629--1647 (2005), preprint at
\url{arxiv.org/abs/physics/0607090}.
    
\bibitem[Schiller 2008]{cs2} See the chapter \emph{General relativity,
deduced from strands}, in C. Schiller, \emph{A Speculation on Unification},
downloadable at \url{www.motionmountain.net}.
    
\bibitem[Schiller 2008b]{cs3} See the chapter \emph{Quantum theory of matter
deduced from strands}, in C. Schiller, \emph{A Speculation on Unification},
downloadable at \url{www.motionmountain.net}.
  
\bibitem[Schiller 2008c]{cs4} See the chapter \emph{Gauge interactions
deduced from strands}, in C. Schiller, \emph{A Speculation on Unification},
downloadable at \url{www.motionmountain.net}.
  
\bibitem[Schiller 2009]{cs5} See the chapter \emph{Particle and their
properties deduced from strands}, in C. Schiller, \emph{A Speculation on
Unification}, downloadable at \url{www.motionmountain.net}.
        
\bibitem[Schwinger 2001]{englert} J. Schwinger, \emph{Quantum Mechanics -- 
Symbolism of Atomic Measurements}, Springer (2001).
        
\bibitem[Smolin \& Wan 2007]{smo} L. Smolin and Y. Wan, \emph{Propagation and
interaction of chiral states in quantum gravity},
\url{arxiv.org/abs/0710.1548}. % and references therein.

\bibitem[Verlinde 2010]{ev} E. Verlinde, \emph{On the origin of gravity and the
laws of Newton}, \url{arxiv.org/abs/1001.0785}.

\bibitem[Wen 2005]{wen} X.-G. Wen, \emph{From new states of matter to a
unification of light and electrons}, \url{arxiv.org/abs/0508020}.

\bibitem[Wilczek \& Zee 1984]{shapwil} F. Wilczek and A. Zee,
\emph{Appearance of gauge structures in simple dynamical systems},
Physical Review Letters {\bf 52}, pp.  2111–2114 (1984),  A.
Shapere and F. Wilczek, \emph{Self-propulsion at low Reynolds number},
Physical Review Letters {\bf 58}, pp.  2051–2054, (1987), and
A. Shapere and F. Wilczek, \emph{Gauge kinematics of deformable
bodies}, American Journal of Physics {\bf 57}, pp.~514–518, (1989).

%-----------------------------------------------------------------------------
%-----------------------------------------------------------------------------
\end{thebibliography}
\end{document}